\documentclass[prx,twocolumn,amssymb,floatfix,showkeywords,amsmath,showpacs]{revtex4}
\usepackage{epsfig}

\usepackage{amssymb}
\usepackage{amsmath}
\usepackage{amsfonts}
\usepackage{bm}
\usepackage{graphicx}
\usepackage{verbatim}
\usepackage{color}
\usepackage[normalem]{ulem}

\newcommand{\bS}{{\bm S}}

\newcommand{\br}{{\bm r}}
\newcommand{\bk}{{\bm k}}

\newcommand{\bR}{{\bm R}}
\newcommand{\bq}{{\bm q}}
\newcommand{\bp}{{\bm p}}

\newcommand{\nn}{{\nonumber}}

\newcommand{\ud}{{\textrm{d}}}

\newcommand{\Ham}{{\mathcal{H}}}

\newcommand{\llangle}{\langle\kern-.25em\langle}
\newcommand{\rrangle}{\rangle\kern-.25em\rangle}

\newcommand{\LLangle}{\Big\langle\kern-.25em\Big\langle}
\newcommand{\RRangle}{\Big\rangle\kern-.25em\Big\rangle}

\begin{document}

\title{Electronic spin-triplet nematic with a twist}

\author{G. Hannappel,$^1$ C.~J. Pedder,$^{1,2}$ F. Kr\"uger,$^{1,3}$ and A.~G. Green$^1$}
\affiliation{$^1$London Centre for Nanotechnology, University College London, Gordon St., London, WC1H 0AH, United Kingdom}
\affiliation{$^2$Physics and Materials Science Research Unit, University of Luxembourg, L-1511 Luxembourg}
\affiliation{$^3$ISIS Facility, Rutherford Appleton Laboratory, Chilton, Didcot, Oxfordshire OX11 0QX, United Kingdom}

\begin{abstract}
We analyze a model of itinerant electrons interacting through a quadrupole density-density repulsion in three dimensions. At the mean 
field level, the interaction drives a continuous Pomeranchuk instability towards $d$-wave, spin-triplet nematic order, which simultaneously 
breaks the SU(2) spin-rotation and spatial rotation symmetries. This order is characterized by spin antisymmetric, elliptical 
deformations of the Fermi surfaces of up and down spins. We show that the effects of quantum fluctuations are similar 
to those in metallic ferromagnets, rendering the nematic transition first-order at low temperatures. Using the fermionic 
quantum order-by-disorder approach to self-consistently calculate fluctuations around possible modulated states,  
we show that the first-order transition is pre-empted by the formation of a helical spin-triplet $d$-density wave. Such a state is closely 
related to $d$-wave bond density wave order in square-lattice systems. 
Moreover, we show that it may coexist with a modulated, $p$-wave superconducting state. 
\end{abstract}

\date{\today}


\maketitle

\section{Introduction}

Electronic liquid crystals are quantum analogues of the classical phases between liquids and solids that partially break
translational and rotational symmetry. For example, in the electron nematic phase, rotational symmetry is broken whilst 
preserving translational invariance. In the two decades since they were proposed \cite{Kivelson+1998}, there has been 
mounting experimental evidence for their existence in a range of systems including cuprate \cite{Ando+2002,Hinkov+2008} 
and pnictide \cite{Chuang+2010,Chu+2012} high-temperature superconductors and two-dimensional electron gases in 
strong magnetic fields \cite{Lilly+1999,Cooper+2002}. 

There are several possible origins for electronic nematicity.  While in cuprates and quantum-Hall systems it could 
be the result of a partial melting of stripe order \cite{Kivelson+1998,Zaanen+2001,Kruger+2002},  in pnictides it may 
be caused by orbital ordering \cite{Kruger+2009,Chen+2009,Lee+2009,Lv+2010,Yin+2010} or else driven by spin-fluctuations 
\cite{Fang+2008,Xu+2008,Eremin+2010,Fernandes+2012}.  The simplest, weak-coupling model consists of an 
interaction in a finite angular momentum channel that drives a distortion of the Fermi surface in that channel 
\cite{Pomeranchuk+1959,Quintanilla+2006,Oganesyan+2001,Kee+2003,Khavkine+2004,Kee+2005,Yamase+2005,Do+2006,
Lawler+2006,Wu+2007,Lawler+2007,Ho+2008,Kee+2009,You+2014}. Whatever its particular microscopic origin, the electron 
nematic supports novel fluctuations and an associated quantum phase transition from the nematic to the conventional Fermi 
liquid \cite{Oganesyan+2001,Lawler+2006, Lawler+2007}. These fluctuations have the potential to drive entirely new 
physics. 

\begin{figure}[t!]
\begin{center}
\includegraphics[width= \linewidth]{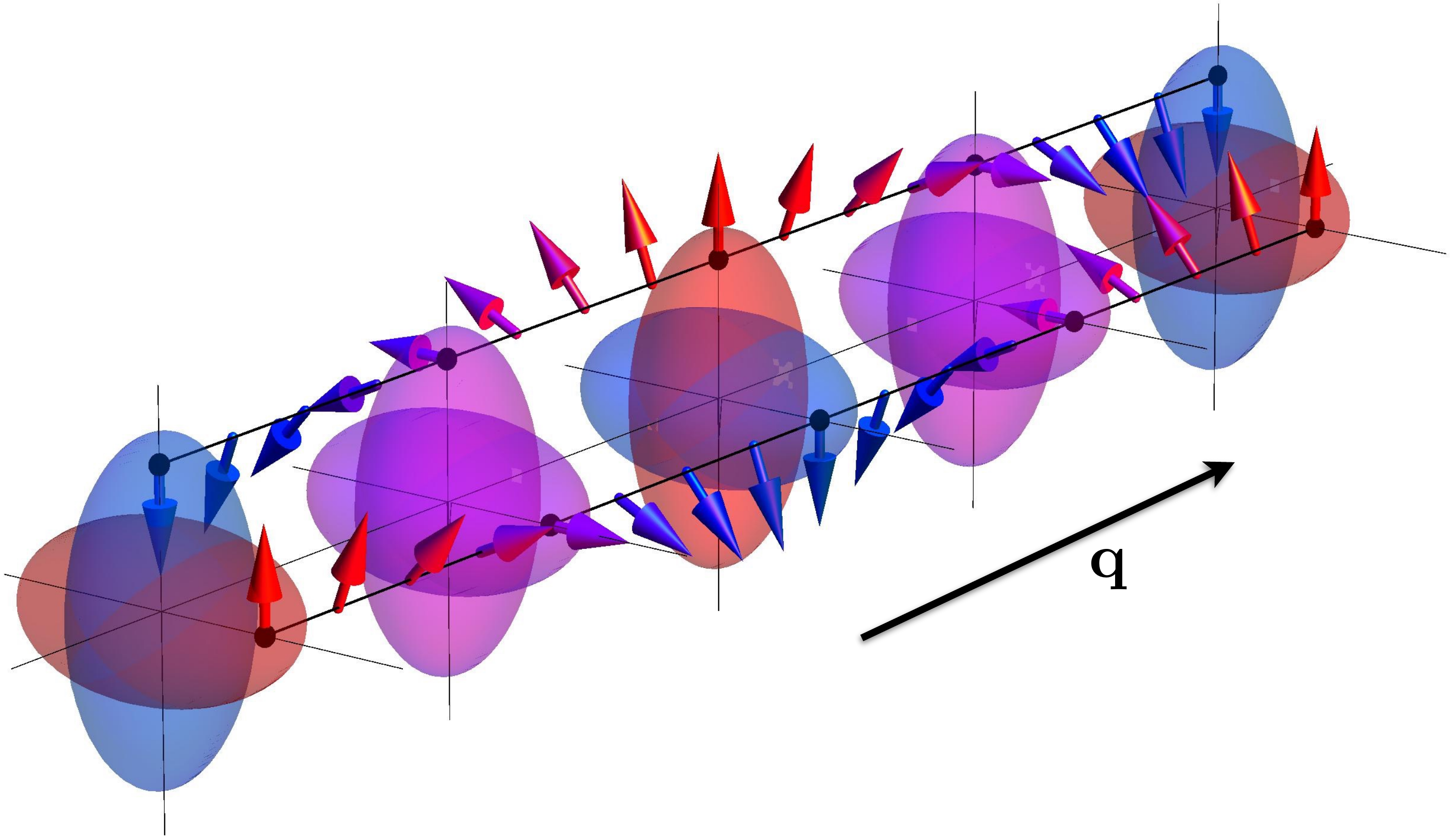}
\caption{The spin-triplet nematic is characterized by spin-antisymmetric elliptical deformations of the Fermi surface,  
simultaneously breaking spatial rotation symmetry and SU(2) spin-rotation symmetry. Fluctuations drive an instability towards the formation of a
$d$-density wave state with a helical modulation in spin space.  Shown is a cartoon of such a state where it is assumed 
that the period of the modulation is much larger than the lattice constant and that Fermi surfaces can be defined on large sub-systems. 
Here, colors represent the spin projection along $z$.}
\label{fig.schematic}
\end{center}
\end{figure}

Strong fluctuations can stabilize new states of collective order. In classical systems, this is an entropic effect. The new state modifies 
the spectrum of fluctuations and thus their entropic contribution to the free energy. Villain's order by disorder picture 
\cite{villain} of frustrated magnets is a transparent realization of this mechanism in which ordered states are 
entropically selected from a degenerate manifold. The central insight -- that the spectrum of fluctuations may ultimately determine the 
state of the system -- finds application further afield, with examples in mechanics \cite{Kapitza} and population dynamics \cite{population}.
It can also be applied to quantum systems. In this case, modification of the spectrum of fluctuations changes their zero-point energy. 
This quantum limit is contained in Villain's model of order-by-disorder for insulators. It is also implicit in the fluctuation induced pairing 
in $^3$He \cite{Balian:1963ve,Anderson+1973,Brinkman:1974} and the ubiquity of new phases near to quantum criticality. 

We study whether the novel fluctuations supported by the spin-triplet electron nematic can drive new collective order. In more familiar 
itinerant ferromagnets, the coupling between Goldstone modes and soft electronic particle-hole fluctuations has profound effects. 
As first shown by Belitz and Kirkpatrick \cite{Belitz+99}, it renders the magnetic phase transition first-order at low temperatures,
as observed in sufficiently clean metallic ferromagnets \cite{Pfleiderer+01,Uemura+07,Otero+08,Taufourr+10,Yelland+11,Shimizu+2015} 
(for a review, see Ref.~\cite{Brando+2015}). Subsequent analysis showed that the first order behavior may be  pre-empted by a spatially 
modulated phase~\cite{Chubukov+04,Conduit+09}, the first clear-cut example of which has been found recently in PrPtAl \cite{AbdulJabbar+2015}. 
It may also be possible for $p$-wave superconductivity to intertwine with this modulated magnetism \cite{Conduit+13}. The close relation 
of these effects to a fermionic version of order-by-disorder was demonstrated in Refs. 
\cite{Duine+05,Conduit+09a,Conduit+09,Karahasanovic+12,Pedder+13,Conduit+13}. 

Hints that similar phenomena might occur in the electron nematic were found in Ref.~\cite{Belitz+2011}, where it was argued that 
 the transition to ``non-$s$-wave ferromagnetism" is driven first-order by fluctuations.  We show that, in fact, fluctuations induce an intertwining 
 of magnetic modulation and $d$-wave nematic order,  resulting in a continuum version of bond density wave order 
 \cite{Metlitski+10a,Metlitski+10b,Efetov+13,Wang+14}. Furthermore, this behavior extends over a larger portion of the phase diagram than the analogous 
 effects in the itinerant ferromagnet. Fluctuations  lead to a co-existent superconducting pairing in the $p$-wave channel,  where the orbital form 
 factor of the superconducting order is  locked to that of the triplet nematic order.  When the twisted ``nematic", triplet $d$-density wave 
 phase meets the superconducting order parameter,  they
 rotate in lockstep, forming a  pair density wave. These unusual phases have some intriguing observable consequences.
For example, whereas the static spin triplet nematic responds to a uniform magnetic field by generating an anisotropic strain \cite{Wu+2007}, 
the  triplet $d$-density wave generates a spatially modulated strain. This offers new possibilities for  experimentally isolating multipolar order 
that may yet prove to be functionally useful. 

The outline of the paper is as follows: In Sec.~\ref{sec.model}, we introduce the electronic model with quadrupole density-density repulsions. 
This model exhibits a spin-triplet nematic ground state for sufficiently strong interactions. In Sec.~\ref{sec.firstorder}, we calculate the fluctuation 
contributions to the free energy and show that they render the nematic transition first-order at low temperatures. Fluctuation-driven instabilities of 
the spin-triplet nematic towards spatial modulation are analyzed in Sec.~\ref{sec.finiteQ}. We show that the first-order transition is pre-empted by 
the formation of a spin-triplet $d$-density wave with a helical modulation of the spin direction.  In Sec.~\ref{sec.pSc}, we study the formation of $p$-wave 
superconductivity in this background. In Sec.~\ref{sec.results}, we calculate the phase diagram and develop an understanding of the homogeneous 
spin-triplet nematic and triplet $d$-density wave  states in both momentum space and real space. Potential observational consequences are discussed in 
Sec.~\ref{sec.exp}. Finally, in Sec.~\ref{sec.disc} we summarize our results and discuss their implications.

\section{Model and Mean Field Theory}
\label{sec.model}

Our starting point is a model of itinerant electrons in three dimensions with isotropic dispersion $\epsilon_0(\bk)\sim k^2$ and a short-ranged  
\emph{quadrupole} density-density repulsion $V(\br)$. At mean-field level, this interaction favors a $d$-wave Pomeranchuk instability in the spin-triplet channel. In momentum 
space, the Hamiltonian can be written as
\begin{eqnarray}
\Ham & = &   \sum_\bk \Psi^\dagger(\bk)\left[\epsilon_0(\bk) -\mu\right] \Psi(\bk)\nn\\
& & +\sum_{\bq,\alpha}  V(\bq) \left[
 \hat{R}_\alpha^s(\bq) \hat{R}_\alpha^s(-\bq)-\hat{\bR}_\alpha^t(\bq)\hat{\bR}_\alpha^t(-\bq)\right],\quad
 \label{eq.ham}
\end{eqnarray}  
where we have adopted the spinor notation, $\Psi=(\psi_\uparrow,\psi_\downarrow)^T$, and defined the quadrupole density operators
\begin{eqnarray}
\hat{R}_\alpha^s(\bq) & = & -\frac{1}{2} \sum_\bk \Psi^\dagger(\bk+\bq)f(\bk) \Phi_\alpha(\bk) \Psi(\bk),\nn \\
\hat{\bR}_\alpha^t(\bq) & = & -\frac{1}{2} \sum_\bk \Psi^\dagger(\bk+\bq) \bm{\sigma} f(\bk) \Phi_\alpha(\bk) \Psi(\bk),\nn
\end{eqnarray}
in the spin singlet ($s$) and triplet ($t$) channels, respectively. This decoupling of the quadrupole density-density repulsion is analogous 
to the conventional splitting of the Coulomb interaction into charge and spin contributions, $\hat{n}_\uparrow\hat{n}_\downarrow = \hat{\rho}^2-\hat{\bS}^2$. 
Note that $\hat{\bR}_\alpha^t(\bk)$ is a three-dimensional vector in spin space [$\bm{\sigma}=(\sigma^x,\sigma^y,\sigma^z)^T$ 
denotes a vector of Pauli matrices]. The additional directional dependence enters through the $d$-wave ($\ell=2$) form factors
$f(\bk) \Phi_\alpha(\bk)$.   In the standard basis,  
$\Phi_1(\bk)=k_x^2-k_y^2$, $\Phi_2(\bk)=(2k_z^2-k_x^2-k_y^2)/\sqrt{3}$, $\Phi_3(\bk)=2k_x k_y$, $\Phi_4(\bk)=2k_x k_z$, 
and $\Phi_5(\bk)=2k_y k_z$. 
 In the definition of the orbital form factors, it is crucial to include a function $f(\bk)$ that is sufficiently peaked at the 
Fermi surface \cite{Chubukov+09}. Otherwise, the neglect of lattice effects and the conventional Coulomb repulsion in our effective low-energy model (\ref{eq.ham}) 
would lead to pathologies such as a divergent electronic density for large nematic order parameters.

The interaction in the nematic channel may have a variety of origins \cite{Oganesyan+2001}. We include it as a phenomenological interaction driving spin 
triplet-nematic order. Following Refs.~\cite{Oganesyan+2001,Kee+2003,Wu+2007}, 
we assume a simple Lorentzian form, 
\begin{equation}
V(\bq)=\frac{g}{1+\xi^2 q^2},
\label{eq.int}
\end{equation}
where $\xi$ parametrizes the range of the interaction.  
For sufficiently strong repulsive interactions $g$, one component of $\hat{\bR}_\alpha^t$ acquires a finite expectation value, $\eta$, corresponding to $d$-wave 
Fermi surface deformations of opposite sign for the two spin species.  This is the same mechanism as the Stoner mean-field theory of ferromagnetism, 
albeit with an extra angular dependence. 
 
\begin{figure}[t!]
\begin{center}
\includegraphics[width= \linewidth]{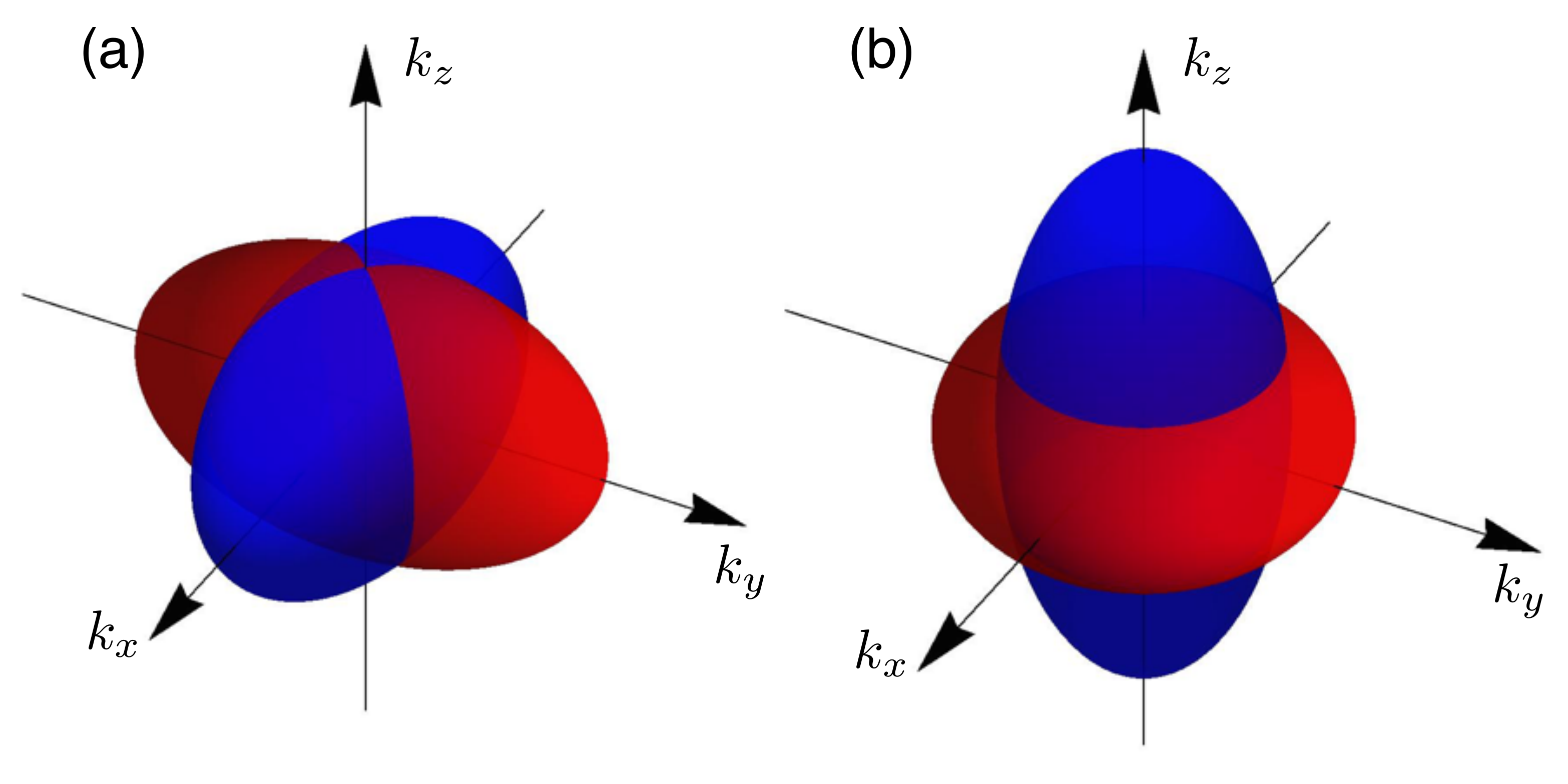}
\caption{Fermi surfaces $A_\uparrow$ (red) and $A_\downarrow$ (blue) of spin-triplet nematic 
states with $e_g$-type $d$-wave deformations $\Phi_1=k_x^2-k_y^2$ (a) and $\Phi_2=(2k_z^2-k_x^2-k_y^2)/\sqrt{3}$ (b).}
\label{fig.dist}
\end{center}
\end{figure}

Since the Hamiltonian (\ref{eq.ham}) does not break spin-rotation symmetry, without loss of generality we choose the $z$-direction as the spin quantization axis. 
In the absence of spatial anisotropy, all of the orbital channels are equivalent, although this degeneracy is broken in any real material by crystal-field anisotropies. 
Throughout the following, we assume that the nematic order develops in the $\alpha=1$ channel. The spin-triplet nematic order parameter is then given by 
$\eta=\langle \hat{R}_1^{z}(\bq=0) \rangle$  and electron dispersion in the presence of this order is
\begin{equation}
\epsilon_\nu(\bk)=k^2-\nu  g \eta f(\bk)\Phi_1(\bk),\nn
\label{eq.disp}
\end{equation}
with  $g=V(0)$.  The resulting mean-field approximation to the free energy at temperature, $T$, is given by 
\begin{equation}
F_\textrm{mf} = g \eta^2 -T\sum_{\nu=\pm 1} \int_\bk \ln \left(e^{-(\epsilon_\nu(\bk)-\mu)/T}+1  \right).
\label{eq.mf}
\end{equation}
Performing a Landau expansion in powers of $\eta$ and absorbing a factor of $g$ into the definition of $\eta$, we obtain
\begin{equation}
F_\textrm{mf}^{(0)} (\eta)= (g^{-1} +\beta_2)\eta^2+\beta_4\eta^4+\beta_6\eta^6.
\label{eq.mf0}
\end{equation}

 In the integrals $\beta_{2n}$ factors of $f(\bk)\Phi_1(\bk)$ occur alongside derivatives of Fermi functions. The latter are strongly peaked 
at the Fermi surface. We can therefore evaluate the orbital form factors at $k_F$ \cite{Wu+2007,Chubukov+09}, 
$f(\bk)\Phi_1(\bk)\to \Phi_1(\hat{\bk})$. In this approximation, the coefficients in the Landau 
expansion are products $\beta_{2n} =  \llangle \Phi_1^{2n}(\hat{\bk}) \rrangle \alpha_{2n}$
of angular averages $\llangle.\rrangle$ over powers of $\Phi_1(\hat{\bk})$ and radially symmetric integrals
\begin{equation}
\alpha_{2n}  =  \frac{1}{n(2n-1)!} \int_\bk  n_F^{(2n-1)}(k^2),\nn
\end{equation}
which are equal to the coefficients in the Landau expansion of the Stoner ferromagnet.
Explicit expressions for the angular averages are derived in Appendix \ref{appendix_angular}. At the mean field level, there is a continuous 
phase transition into a spin-triplet nematic state, determined by the condition $g\beta_2(T)=-1$. 

As a side remark, we note that spin-triplet Pomeranchuk instabilities can also occur without rotational symmetry breaking in real space \cite{Wu+2004,Wu+2007}. 
By analogy with A and B phases of superfluid helium-3 \cite{Leggett1975}, one distinguishes between $\alpha$ and $\beta$ phases of spin-triplet 
Pomeranchuk systems. The $\alpha$ phases are characterized by spin-antisymmetric Fermi-surface deformations as discussed above. The $\beta$ phases 
retain the symmetry of the undistorted Fermi surface but exhibit vortex structures in momentum space with winding numbers $\pm\ell$ \cite{Wu+2007}.
In this work we do not consider such $\beta$ phases.

\section{Fluctuation Contributions to Free Energy of the Spin-Triplet Nematic}
\label{sec.fluct}
The central result of this paper is the prediction of new phases that are driven by fluctuations near to the spin-triplet nematic quantum critical point. 
It has already been argued in Ref.~\cite{Belitz+2011} that any Pomeranchuk instability in the spin-triplet channel will ultimately be driven first-order by fluctuations. In the 
related itinerant ferromagnet,  these same fluctuations are responsible for a much richer set of instabilities,  so the appearance of novel phases driven by nematic fluctuations 
is to be expected. Physically, the instabilities are driven by the interplay of the Goldstone modes with soft particle-hole excitations. This leads to non-analyticities in the 
Ginzburg-Landau expansion. Alternatively, new phases constructed within the background of spin-triplet nematic order modify the spectrum 
of fluctuations and so modify the zero-point energy. This  change of the free energy landscape sequentially drives 
the spin-triplet nematic transition first order,  then to develop spatial modulations and coexistent, $p$-wave superconductivity.

\subsection{Fluctuation-Driven First Order Transition}
\label{sec.firstorder}
We begin by investigating how fluctuations modify the transition into a phase of uniform $d$-wave spin-triplet nematic order. These effects can be accommodated 
diagrammatically -- as has been demonstrated for the $p$-wave ($\ell=1$), spin-triplet Pomeranchuk instability \cite{Belitz+2011}. Here, however, we self-consistently 
calculate fluctuations around the ordered, broken-symmetry state, using the fermionic quantum order-by-disorder approach. This  technique reveals the underlying physics 
more directly.  For the itinerant ferromagnet in three dimensions, this procedure  reproduces the diagrammatic result $F_\textrm{fl}(m)\sim m^4\ln(m^2+T^2)$, on the level of 
self-consistent second-order perturbation theory \cite{Pedder+13}.  

Because of the angular dependence of the orbital form factors, $\Phi_\alpha(\bk)$,  the non-analyticities have a different form compared to those of the itinerant ferromagnet. 
This is important for the phase behavior as $T\to0$ and for the instabilities of the spin-triplet nematic towards spatially modulated order. 
The behavior for small values of the order parameter is, however, essentially the same as that of the ferromagnet. Specifically, we find the same $\ln T$ contribution to the $\eta^4$ 
coefficient, which is responsible for the first-order transition at low temperatures. 

The details of the calculation are very similar to those in the ferromagnetic case. We first express the partition function as an imaginary-time path integral over fermionic 
 fields $\Psi(\br,\tau)=[\psi_\uparrow(\br,\tau),\psi_\downarrow(\br,\tau)]^T$, and then decouple the quadrupole interaction (\ref{eq.ham}) by a Hubbard-Statonovich transformation, 
\begin{eqnarray}
\mathcal{S}_\textrm{int} & = & \int_\tau \sum_{\bq,\alpha} V(\bq)\Big\{|\bm{\phi}_\alpha(\bq,\tau)|^2 - |\rho_\alpha(\bq,\tau)|^2 \nn\\
&  &  + \sum_\bk \Psi^\dagger (\bk+\bq,\tau)\left[\rho_\alpha(\bq,\tau)-\bm{\phi}_\alpha(\bq,\tau)\cdot\bm{\sigma}\right]\nn\\
& & \quad\quad \times f(\bk)\Phi_\alpha(\bk) \Psi(\bk,\tau)\Big\}.\nn
\end{eqnarray}
The twenty fields $\rho_\alpha$ and $\bm{\phi}_\alpha$ correspond to a single spin-symmetric and three spin-antisymmetric fluctuations, respectively,  in each of the five orbital 
channels. The spin-triplet nematic order parameters, $\eta^i_\alpha$, are given by the zero-frequency components of the 
fluctuation fields, $\phi^i_\alpha(\br,\omega)=\eta^i_\alpha+\tilde{\phi}^i_\alpha(\br,\omega)$ with $\tilde{\phi}^i_\alpha(\br,\omega=0)=0$. As previously described, we consider 
elliptical Fermi surface distortions in the $\alpha=1$, $(k_x^2-k_y^2)$ channel.

In order to facilitate the self-consistent free energy expansion, we include the static nematic order parameter $\eta$ in the free-fermion action, 
\begin{eqnarray}
\mathcal{S}_0[\overline{\Psi},\Psi,\eta] 
& = &  
\sum_{\nu=\pm1} 
\sum_{\bk, \omega}
\overline{\psi}_\nu(\bk,\omega) G^{-1}_\nu(\bk,\omega)\psi_\nu (\bk,\omega),
\nn\\
G_\nu(\bk,\omega) 
& = & 
\frac{1}{-i\omega+k^2-\nu g \eta f(\bk) \Phi_1(\bk)-\mu},\nn
\end{eqnarray}
where $g=V(0)$, and we redefine the interaction in terms of the finite-frequency parts of the fluctuation fields, 
$\mathcal{S}_\textrm{int}[\overline{\Psi},\Psi,\tilde{\rho}_\alpha,\tilde{\bm{\phi}}_\alpha]$ only. The free energy can in principle be expressed as a functional of this Green's 
function - the Kadanoff-Baym approach \cite{Kadanoff+1962} - or equivalently viewed as a functional of the mean field dispersion. 

The next steps involve integrating over the fermionic fields, followed by expanding in fluctuation fields up to quadratic order and integrating over them. The result is
\begin{eqnarray}
F_\textrm{fl}  & = &  -\frac{T}{2} \sum_{q,\tilde{\omega}}\sum_{\alpha,\beta} V^2(\bq)\left[  \Pi_{++}^{\alpha\beta}(\bq,\tilde{\omega}) \Pi_{--}^{\alpha\beta}(\bq,\tilde{\omega})\right.\nn\\
& & \quad\quad +\left.\Pi_{+-}^{\alpha\beta}(\bq,\tilde{\omega}) \Pi_{-+}^{\alpha\beta}(\bq,\tilde{\omega})\right],\nn
\end{eqnarray}
with $\tilde{\omega}$ a bosonic Matsubara frequency. We have defined 
\begin{eqnarray}
\Pi_{\nu,\nu'}^{\alpha\beta} (\bq,\tilde{\omega}) 
& = & 
T\sum_{\bk,\omega}
G_\nu(\bk,\omega)G_{\nu'}(\bk+\bq,\omega+\tilde{\omega})
\nn\\
& & \quad\quad \times 
f(\bk)\Phi_\alpha(\bk) f(\bk+\bq)\Phi_\beta(\bk+\bq).\nn
\end{eqnarray}
After summation over Matsubara frequencies, we obtain
\begin{eqnarray}
F_\textrm{fl} 
& = & 
 \frac{1}{2} \sum_{\bk_1,\ldots,\bk_4}
  \delta_{\bk_1-\bk_2,\bk_3-\bk_4}V^2(\bk_1-\bk_2) 
  \Omega(\bk_1,\ldots,\bk_4)
  \nn\\
& & \times 
\frac{ n_F({\epsilon}_{\bk_1}^+) n_F({\epsilon}_{\bk_2}^-) 
[n_F({\epsilon}_{\bk_3}^+) +n_F({\epsilon}_{\bk_4}^-)]}{\epsilon_{\bk_1}^+ +\epsilon_{\bk_2}^- -\epsilon_{\bk_3}^+ -\epsilon_{\bk_4}^-},
\label{eq.fl0}
\end{eqnarray}
where 
$\Omega(\bk_1,\ldots,\bk_4) \! =\!  \sum_{\alpha,\beta}\Phi_\alpha(\hat \bk_1) \Phi_\beta(\hat \bk_2) \Phi_\alpha(\hat \bk_3) \Phi_\beta(\hat \bk_4)$. 
Note that as in the mean-field calculation, we evaluate the orbital form factors at the Fermi surface. At small temperatures, the main contribution to the fluctuation 
integral (\ref{eq.fl0}) comes from momenta that are (anti-)parallel and close to the Fermi wave-vector. We can therefore approximate $V(|\bk_1-\bk_2|)\approx V(2k_F)$
and  $\Omega(\bk_1,\ldots,\bk_4) \approx  (\sum_{\alpha}\Phi_\alpha^2(\hat{\bk}) )^2=16/9$, which is fixed by the normalization of the spherical harmonics. 
A similar summation results for instabilities in higher angular momentum channels. 

After this approximation, Eq.~(\ref{eq.fl0}) has exactly the same form as in the ferromagnetic case~\cite{Karahasanovic+12}, but with the magnetization, $m$, replaced by $\eta \Phi_1(\hat \bk)$. Expanding in powers of $\eta$, the coefficient of the $\eta^{2n}$ term is proportional to the $m^{2n}$ coefficient for the ferromagnet, with a proportionality factor that is given by an angular integral over $\Phi_1^{2n}(\hat{\bk})$. 
As a result of these considerations, we can obtain the fluctuation contribution to the free energy of the spin-triplet nematic from an angular average of the ferromagnetic result. Using the result of Ref. \cite{Pedder+13}, which re-sums leading temperature divergences to all orders in the magnetization, we obtain
\begin{eqnarray}
F_\textrm{fl} (\eta) &  = & c V^2(2 k_F) \LLangle -2(1+2\ln2) \Phi_1^2(\hat \bk) \eta^2+2 \Phi_1^4(\hat \bk)  \eta^4 \nn\\
& & \quad\quad+  \Phi_1^4(\hat \bk) \eta^4\ln\frac{\kappa^2   \Phi_1^2(\hat \bk)\eta^2+T^2}{\mu^2} \RRangle,
\label{eq.fl}
\end{eqnarray}
where, as in the mean-field free energy, we have rescaled $\eta$ to include a factor of $g$. Furthermore, $c=\frac{16}{9} c_\textrm{FM}= \frac{16}{9}\cdot\frac{8\sqrt{2}}{3(2\pi)^6}$, 
and $\kappa$ is a phenomenological parameter that accounts for the renormalization due to sub-leading fluctuation corrections \cite{Pedder+13}.
In terms of the angular averages $\llangle \Phi_1^{2n}(\hat {\bf k}) \rrangle$ [see Appendix \ref{appendix_angular}] the fluctuation contribution can be rewritten as
 \begin{eqnarray}
\frac{F_\textrm{fl}(\eta)}{c V^2(2 k_F)} 
&=&  -2(1+2\ln2) \llangle  \Phi_1^2(\hat {\bk}) \rrangle \eta^2 \nn \\ 
& & + 2[1+\ln(T/\mu)] \llangle  \Phi_1^4(\hat {\bk}) \rrangle \eta^4\nn\\
& & + \Omega_0(\kappa^2\eta^2/T^2)\eta^4\nn\\
\Omega_0(x) &=&  \sum_{k=1}^\infty \frac{(-1)^{k-1}}{k} 
\llangle \Phi_1^{2(k+2)}(\hat {\bk}) \rrangle  x^k.
\label{eq.fluc0}
 \end{eqnarray}
As for the ferromagnet, fluctuations give rise to a $\ln T$ contribution to the $\eta^4$ coefficient, causing the transitions to become first-order at sufficiently low temperatures.  
The tri-critical point below which the transition is discontinuous is determined by the simultaneous vanishing of the full coefficients of  $\eta^2$ and $\eta^4$,
\begin{eqnarray}
0 & = & g^{-1} +\beta_2(T)-2(1+2\ln2) cV^2(2 k_F) \llangle  \Phi_1^2(\hat {\bk}) \rrangle, \nn  \\
0 & = & \beta_4(T)+2 cV^2(2 k_F)  [1+\ln(T/\mu)]\llangle \Phi_1^4(\hat {\bk})\rrangle.
\label{eq.tricrit}
\end{eqnarray}
The function $\Omega_0(x)$ does not affect the location of the tri-critical point --- it is a special hypergeometric function that is positive definite for $x\ge 0$ and vanishes linearly as $x\searrow 0$. The resulting contributions are, therefore, at least of order $\eta^6$. The behavior of $\Omega_0(x)$ is crucial, however, for the phase stability at temperatures below the tri-critical point.

\subsection{Finite $\bq$ instability}
\label{sec.finiteQ}
Fluctuation induced first-order behavior  often heralds instabilities towards other, competing order. For example, itinerant ferromagnets are unstable to modulated  magnetic or 
helimagnetic order below the tri-critical point, where fluctuations drive the phase transition first-order. 

For the spin-triplet nematic driven by quadrupole interactions, the similarity of the fluctuation corrections (\ref{eq.fluc0}) suggests that the  free energy could be lowered by the formation of modulated 
``nematic", triplet $d$-density wave order. Here the situation is much richer since the order parameter, $\bm{\eta}(\br) = \langle  \hat{\bR}^t(\br) \rangle$, is a 15-dimensional vector in spin-orbital product space. Modulation may consist of rotation between any of its components. The possibilities are reduced by allowing for physical effects in materials. Firstly, the modulation must couple to the electron spin in order to be favored by spin fluctuations. We also expect that modulation between different orbital components is suppressed by crystal field anisotropy. Allowing for these considerations, we investigate helical spin-triplet $d$-density wave order as indicated in Fig.~\ref{fig.schematic}. This consists of a rotation of the spin quantization axis in the $xy$-plane (for example) with a pitch $\bq$. Its order parameter is given by
\begin {eqnarray}
\eta & = & - \frac{1}{2}  \sum_\bk \left[ \langle \Psi^\dagger(\bk+\bq) \sigma^+ f(\bk)\Phi_1(\bk) \Psi(\bk) \rangle\right.\nn\\
& &\quad\quad+\left.\langle \Psi^\dagger(\bk-\bq) \sigma^- f(\bk)\Phi_1(\bk) \Psi(\bk)\rangle\right].\nn
\label{eq.helical_order}
\end{eqnarray}

In the absence of any Fermi-surface nesting, such a modulated state is certainly not favored by a Pomeranchuk mean-field instability. It can be favored by fluctuations. In order to show this, we again use the fermionic quantum order-by-disorder approach, extending it to self-consistently calculate the fluctuations around the broken-symmetry states, characterized by the order parameter $\bm{\eta}(\br)$.
Since the spin-triplet nematic order breaks the spatial rotation symmetry, the free energy depends upon the direction of $\bq$. For $\bq=0$, the order parameter reduces to that of the homogeneous spin-triplet nematic, while for $\eta=0$ we obtain the disordered metallic state with isotropic and identical Fermi-surfaces of spin-up and spin-down electrons.  

The calculation proceeds in the same manner as that for the uniform spin-triplet nematic described in Sec. \ref{sec.fluct}. By self-consistently calculating the mean-field and fluctuation contributions, we can express the free energy as a functional of the mean-field electronic dispersion in the presence of modulated order $\bm{\eta}(\br)$.

The mean-field Hamiltonian in the presence of helical triplet $d$-density wave order
is easily diagonalized by a transformation to the rotating frame, 
\begin{equation}
\left( \begin{array}{c} \varphi_+(\bk) \\  \varphi_-(\bk)  \end{array}  \right) = e^{i\frac{\theta(\bk)}{2}\sigma_y}
\left( \begin{array}{c} \psi_\uparrow(\bk+\bq/2) \\  \psi_\downarrow(\bk-\bq/2)  \end{array}  \right),
\label{eq.diagonal}
\end{equation}
with $\tan \theta(\bk)= g\eta  f(\bk)\Phi_1(\bk)/(\bk.\bq)$, yielding the electron dispersion 
\begin{equation}
\epsilon_\nu(\bk) = k^2-\nu\sqrt{(\bk.\bq)^2+g^2\eta^2  f^2(\bk)\Phi_1^2(\bk)}.
\label{nematicdisp}
\end{equation} 
The mean field free energy is obtained by inserting the dispersion, Eq.~(\ref{nematicdisp}), into Eq\sout{s}.~(\ref{eq.mf}). After expanding in powers of $\eta$  and $\bq$, and 
constraining the order-parameter coupling to the vicinity of the Fermi surface as before,  we obtain 
\begin{eqnarray}
F_\textrm{mf}(\eta,\bq) & = &  F_\textrm{mf}^{(0)}(\eta) +2\alpha_4 \llangle \Phi_1^2(\hat {\bk})( \hat {\bk}.\hat {\bq} )^2  \rrangle  \eta^2 q^2\nn\\
& & +3\alpha_6 \llangle \Phi_1^4(\hat {\bk})  ( \hat {\bk}.\hat {\bq} )^2 \rrangle  \eta^4q^2  \nn\\
& & + 3\alpha_6 \llangle  \Phi_1^2(\hat {\bk}) ( \hat {\bk}.\hat {\bq} )^4 \rrangle \eta^2 q^4,
\label{eq.FEmfq}
\end{eqnarray}
where $F_\textrm{mf}^{(0)}(\eta)$ denotes the mean-field free energy of the homogeneous spin-triplet nematic (\ref{eq.mf0}). We have again absorbed a factor of $g$ in the 
definition of $\eta$. Explicit expressions for the angular averages are computed in Appendix A for high-symmetry directions of $\bq$. Since the angular averages and the 
coefficients $\alpha_4$ and $\alpha_6$ are always positive, spatial modulations  lead to an increase of $F_\textrm{mf}$ and are therefore not favored at mean-field. 

To self-consistently calculate the fluctuation contributions, we include the modulated order parameter in the free-fermion propagator, which after diagonalization (\ref{eq.diagonal}) becomes 
$G_\nu(\bk,\omega) =[-i\omega+\epsilon_\nu(\bk)-\mu]^{-1}$, 
with the electron dispersion given in Eq.~(\ref{nematicdisp}). Transforming the finite-frequency fluctuation fields to the rotating frame, we can proceed in exactly the same way as in the homogeneous case (Sec.~\ref{sec.firstorder}). The resulting free-energy contributions are obtained by replacing $\eta \Phi_1(\hat {\bk})$  in Eq.~(\ref{eq.fl}) with $\sqrt{( \hat {\bk}.\bq )^2+\eta^2\Phi_1^2(\hat{\bk})}$. Taking account of the angular averages, we obtain $F_\textrm{fl}(\eta,\bq) = F_\textrm{fl}^{(0)}(\eta)+\delta F_\textrm{fl}(\eta,\bq)$ with
\begin{eqnarray}
 \label{eq.FEflucq}
\frac{\delta F_\textrm{fl}(\eta,\bq)}{cV^2(2k_F)} & = & 4 [1+\ln(T/\mu)] \llangle  \Phi_1^2( \hat{\bk}) ( \hat{\bk}.\hat{\bq} )^2 \rrangle \eta^2 q^2 \\
& & +  \Omega^{\hat{\bq}}_2(\kappa^2\eta^2/T^2) \eta^2 q^2
 + \Omega^{\hat{\bq}}_4(\kappa^2\eta^2/T^2)q^4, \nn
\end{eqnarray}
where the functions 
\begin{eqnarray}
\Omega^{\hat{\bq}}_2(x) & = & \sum_{k=1}^\infty \frac{(-1)^{k-1}}{k}\binom{k+2}{1} \llangle \Phi_1^{2(2k+1)}(\hat{\bk})  ( \hat{\bk}.\hat{\bq} )^2  \rrangle  x^k, \nn \\
\Omega^{\hat{\bq}}_4(x) & = & \sum_{k=1}^\infty \frac{(-1)^{k-1}}{k}\binom{k+2}{2}  \llangle  \Phi_1^{2k}(\hat{\bk}) ( \hat{\bk}.\hat{\bq} )^4 \rrangle x^k
\label{eq.Omegaq}
\end{eqnarray}
are positive for $x>0$ and vanish linearly as $x\searrow 0$. This result shows that the coupling to soft electronic particle-hole fluctuations gives rise to  $\ln T$ dependence of the $\eta^2 q^2$ and $\eta^4$ coefficients. In fact, the coefficients are strictly proportional to each other, with a proportionality factor that is independent of temperature and the same for mean-field and fluctuation contributions. As a result, the coefficients change sign at the same temperature, and so the first order transition into the homogeneous spin-triplet nematic state is pre-empted by the formation of a modulated  triplet $d$-density wave  state. The direction of the modulation vector $\bq$ depends upon the angular averages  and the behavior of the functions 
$\Omega^{\hat{\bq}}_2$ and $\Omega^{\hat{\bq}}_4$ (\ref{eq.Omegaq}).

\subsection{Superconducting Instability}
\label{sec.pSc}

That magnetic fluctuations mediate the formation of Cooper pairs was first realized in the context of superfluid $^3$He \cite{Balian:1963ve,Anderson+1973,Brinkman:1974}. The translation of 
this idea to itinerant ferromagnets and the potential instability to $p$-wave superconductivity 
 was first suggested in Ref.~\cite{Fay:1980kx}. Given the similarities between the physics of itinerant ferromagnets and spin-triplet nematics, one could wonder whether the spin-triplet nematic has a similar instability to superconductivity. Indeed, the discovery of the pnictide superconductors with their nematic order and superconducting transition has made this a very active line of investigation. 

We follow Ref.~\cite{Conduit+13} and use the fermionic order-by-disorder approach to investigate the possibility of $p$-wave superconductivity in the spin-triplet nematic. We find that fluctuations in this phase do indeed drive superconductivity. The state displays a subtle interplay of superconducting and spin-triplet nematic order parameters --- the orientation of their orbital form factors being locked together. In the region of the phase diagram where  fluctuations stabilize a spatially modulated, triplet $d$-density wave, the intertwining with $p$-wave superconductivity leads to an entirely new phase that we will discuss later on.

There are two ways in which one might incorporate superconducting instabilities into the fermionic order-by-disorder approach. The first is via a Legendre transform in which one introduces a field conjugate to the superconducting order parameter. The quadratic parts of the Hamiltonian are diagonalized and  the interacting parts treated using the fermionic order by disorder approach -- including spin-triplet nematic order. The resulting generating functional for superconductivity in the presence of the spin-triplet nematic is Legendre transformed back to obtain a free energy function. This is similar in spirit to using the density functional for superconductivity recently introduced by Hardy {\it et al.} \cite{Hardyetal2014} to describe spin fluctuation-induced superconductivity \cite{KB}.

The alternative approach, which is equivalent for continuous transitions, is to make a variational ansatz \cite{Conduit+13}. The general scheme is as follows: (i) after having first diagonalized the 
electron state in the spin-nematic background, we add and subtract a term 
$\delta{\cal H}(\Delta)=\sum_{\bk} \left( \Delta \varphi^\dagger_{-{\bk},+}  \varphi^\dagger_{{\bk},+} +\textrm{h.c.} \right)$
in the Hamiltonian, where $\varphi^\dagger_{{\bk}+}$ creates an electronic state that is diagonal in the presence of  spin-triplet nematic or $d$-density wave order, Eq.~(\ref{eq.diagonal}). 
(ii) The quadratic terms ${\cal H}_0 -\delta{\cal H}(\Delta)$ can be diagonalized by a Bogoliubov transformation. (iii) The remaining terms 
${\cal H}_\textrm{int} +\delta{\cal H}(\Delta)$
can be treated using the fermionic order-by-disorder approach, accounting for the change of interaction vertex imposed by the Bogoliubov transformation. 
Expanding to quadratic order in the superconducting order parameter, we obtain
\begin{eqnarray}
F_\textrm{SC}(\Delta) & = & -\sum_{\bk}  \frac{ 2 n^\uparrow _{\bk} -1 }{2 \xi^\uparrow_{\bk} } \left[ 1 - \partial_{\epsilon_{\bk}} {\cal R}e \Sigma^\uparrow ({\bk}, \epsilon_{\bk} ) \right]| \Delta_{\bk}|^2\nn\\
& & + g^2 \sum_{{\bk},{\bq} } 
\frac{ 2 n^\uparrow _{{\bk}+{\bq}} -1 }{2 \xi^\uparrow_{{\bk}+{\bq}} }    \bar \Delta_{{\bk}+{\bq}} \frac{ 2 n^\uparrow _{\bk} -1 }{2 \xi^\uparrow_{\bk} }      \Delta_{{\bk}} \nn\\
& & \quad\quad\quad\times {\cal R}e \chi^{\downarrow \downarrow} ({\bq}, \epsilon_{{\bk}+{\bq}}^\uparrow -\epsilon_{{\bk}}^\uparrow),\nn
\label{ScF}
\end{eqnarray}
as additional contributions to the Ginzburg-Landau free energy, where $\xi_{\bk}^\nu = \epsilon_\nu(\bk) - \mu$ with $ \epsilon_\nu(\bk)$ given by Eq.~(\ref{nematicdisp}).
$\chi$ and $\Sigma$ are the magnetic susceptibility and self-energy evaluated in the presence of  spin-triplet nematic or $d$-density wave order.
They are calculated explicitly in Appendix \ref{appendix_Scangular}. This is similar to the additional contributions found in the case of $p$-wave instabilities of the itinerant 
ferromagnet \cite{Conduit+13}. Indeed, the only differences are some additional angular factors arising from the form factors of the nematic order.

In order to determine the superconducting transition temperature, we assume that the superconducting pairing occurs only very near the Fermi surface. We take $\Delta_{\bk}= \Delta \Theta_{\bk}$, 
where $\Theta_{\bk}$ is the orbital form factor of the $p$-wave superconducting order. In addition, we approximate the factors of $(2 n^\uparrow _{\bk} -1 )/(2 \xi^\uparrow_{\bk} )$ by delta functions 
at the Fermi surface weighted by a suitable pre-factor,
$(2 n^\uparrow _{\bk} -1 )/(2 \xi^\uparrow_{\bk} )
\approx
\chi_\Delta^0 \delta (\epsilon_{\bk}^\uparrow - \mu )$, 
with $\chi^0_\Delta=\ln \left[ ( 2 \mu e^{\cal C})/( \pi T) \right]$ the bare susceptibility to superconducting order (and ${\cal C} \approx 0.577$ is the Euler constant). This amounts to an approximation that pairing 
only occurs at the Fermi surface. 

Using these approximations and definitions, the superconducting transition temperature is determined by the vanishing of the quadratic coefficient of $\Delta$. The result is
\begin{eqnarray}
T_\textrm{SC}
&=&
\frac{2 \mu e^{\cal C}}{\pi} 
\exp
\left[
-\frac{
\langle \langle 
\Theta_{{\bk}+{\bq}} \Theta_{\bk}
{\cal R}e 
\chi^{\downarrow \downarrow} ({\bq}, \epsilon_{{\bk}+{\bq}}^\uparrow -\epsilon_{{\bk}}^\uparrow)
\rangle \rangle
}{
\langle \langle 
\Theta_{\bk}^2 
\left[ 1 - \partial_{\epsilon_{\bk}} {\cal R}e \Sigma^\uparrow ({\bk}, \epsilon_{\bk} ) \right]
\rangle \rangle
}
\right],
\nn\\
\label{eq.FayAppel}
\end{eqnarray}
which is the spin-triplet nematic analogue of that obtained by Fay and Appel for the ferromagnet \cite{Fay:1980kx}. In Eq.~(\ref{eq.FayAppel}), 
$\langle \langle ... \rangle \rangle =\sum_{\bk} ...  \, \delta(\epsilon^\uparrow_\bk - \mu)$ or
$\sum_{\bk,\bq} ... \,  \delta(\epsilon^\uparrow_\bk - \mu) \delta(\epsilon^\uparrow_{\bk+\bq} - \mu)$ as appropriate and indicates an average over the  Fermi surface of the pairing electrons.

It is important to note that in order to derive the expression for the superconducting transition temperature $T_\textrm{SC}$ (\ref{eq.FayAppel}), we have made an additional 
approximation. Instead of minimizing the full free energy with superconducting and spin-triplet nematic order parameters, we have analyzed the pairing instability on the background of spin-triplet 
nematic or $d$-density wave order, neglecting any feedback of the superconductivity on this background order. This procedure seems justified since experiments on closely related 
metallic ferromagnets such as UGe$_2$ \cite{Saxena+00} find a coexistence of the magnetic order with $p$-wave superconductivity with only small changes of the magnetization across  $T_\textrm{SC}$. 

In the case of superconducting order in a ferromagnetic background, the self energy and spin susceptibility are uniform over the Fermi surface and the only angular dependence comes from the 
superconducting form factors. The resulting angular integrals can be carried out as in Ref. \cite{Fay:1980kx}. In the present case, the self energy and spin susceptibility depend upon the spin-triplet 
nematic order and so inherit an angular dependence as a result.  This has important consequences. 

\begin{figure}[t!]
\begin{center}
\includegraphics[width= 0.65 \linewidth]{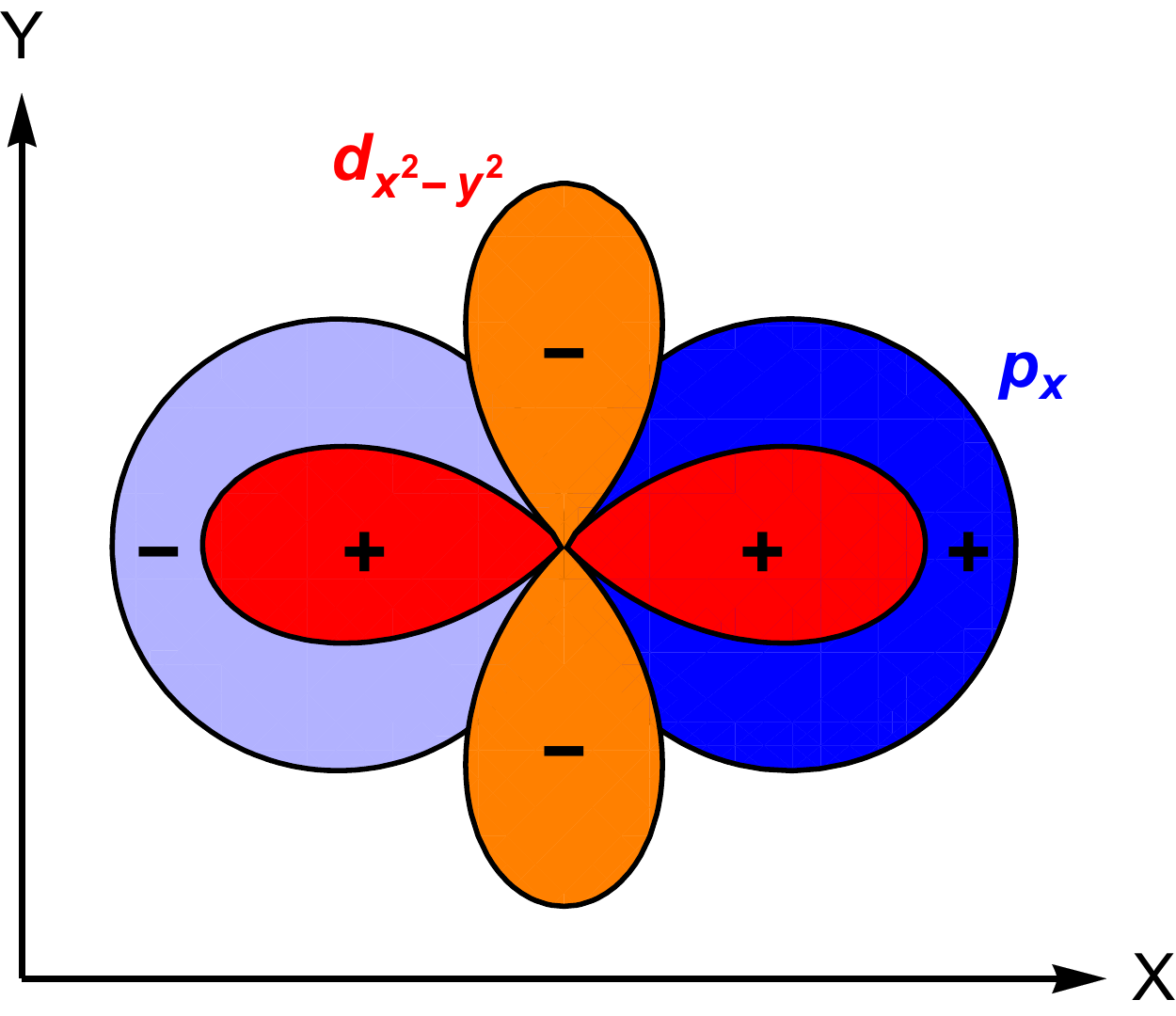}
\caption{Locking of relative orbital orientation of $d$-wave spin-triplet nematic order and $p$-wave superconductivity: mode-mode coupling enhances the superconducting transition temperature for orbital order with the relative orientation shown. Other orientations of the superconducting form factor are disfavored.}
\label{fig.orientation}
\end{center}
\end{figure}

In the ferromagnet, mode-mode coupling effects lead to an enhanced superconducting transition temperature in the ferromagnetically ordered part of the phase diagram compared to the 
paramagnetic part \cite{Roussev2001,Kirkpatrick2001,Wang2001}. A similar effect occurs for $p$-wave superconductivity in the spin-triplet nematic. However, the enhancement occurs only for 
the relative orientation of superconducting form factor $\Theta_\bk$ and spin-triplet nematic form factor $\Phi_\bk$ shown in Fig.~\ref{fig.orientation}. In other relative orientations, the mode-mode coupling 
terms -- which enter the free energy as coefficients of terms of the form $\Delta^2 \eta$ -- are zero or even disfavor $p$-wave superconductivity. This effect pins the relative orientation of the 
orbital components of spin-triplet nematic and superconducting order.

Evaluation of the integrals in Eq.~(\ref{eq.FayAppel}) is tedious and we relegate the details to Appendix \ref{appendix_Scangular}. For weak spin-triplet nematic order, $\eta\ll1$,  we find
\begin{eqnarray}
& &
\langle \langle 
\Theta_{{\bk}+{\bq}} \Theta_{\bk}
{\cal R}e 
\chi^{\downarrow \downarrow} ({\bq}, \epsilon_{{\bk}+{\bq}}^\uparrow -\epsilon_{{\bk}}^\uparrow)
\rangle \rangle\nn
\\
&\approx&
\!\!
\lambda\left[0.026{+}0.084  \eta+  (0.057{-}0.113 \ln T) \eta^2\right]\nn
\\
& &
\langle \langle 
\Theta_{\bk}^2 
\left[ 1 - \partial_{\epsilon_{\bk}} {\cal R}e \Sigma^\uparrow ({\bk}, \epsilon_{\bk} ) \right]
\rangle \rangle\nn
\\
& \approx&
\!\!
\lambda
\left[0.398{+}0.199  \eta{+}\left( 0.976{+}0.060 \ln T \right)  \eta^2 \right]\nn
\end{eqnarray}
with $\lambda=-\frac {16} {9}/(2\pi)^6$ and a factor of $g$ is absorbed in the redefinition of $\eta$, as before. For stronger spin-triplet nematic order, the superconductivity is suppressed --- the 
regions of the Fermi surface with low energy magnetic fluctuations that drive Cooper pairing are reduced in size.

\section{The Phase Diagram of the Helical Spin-Triplet Nematic}
\label{sec.results}

In the preceding sections, we developed a Ginzburg-Landau expansions for the spin-triplet nematic, allowing for formation of helical modulated phases and superconductivity. With this in hand, 
in this Section, we will analyze the phase diagram that results as a function of temperature $T$ and quadrupole interaction strength $g$.  In real materials, the latter can be tunable by doping 
or pressure. Finally, we will develop an understanding of  the helical spin-triplet $d$-density wave  in both momentum space and real space. 

\subsection{Phase Diagram}
\begin{figure}[t!]
\begin{center}
\includegraphics[width= \linewidth]{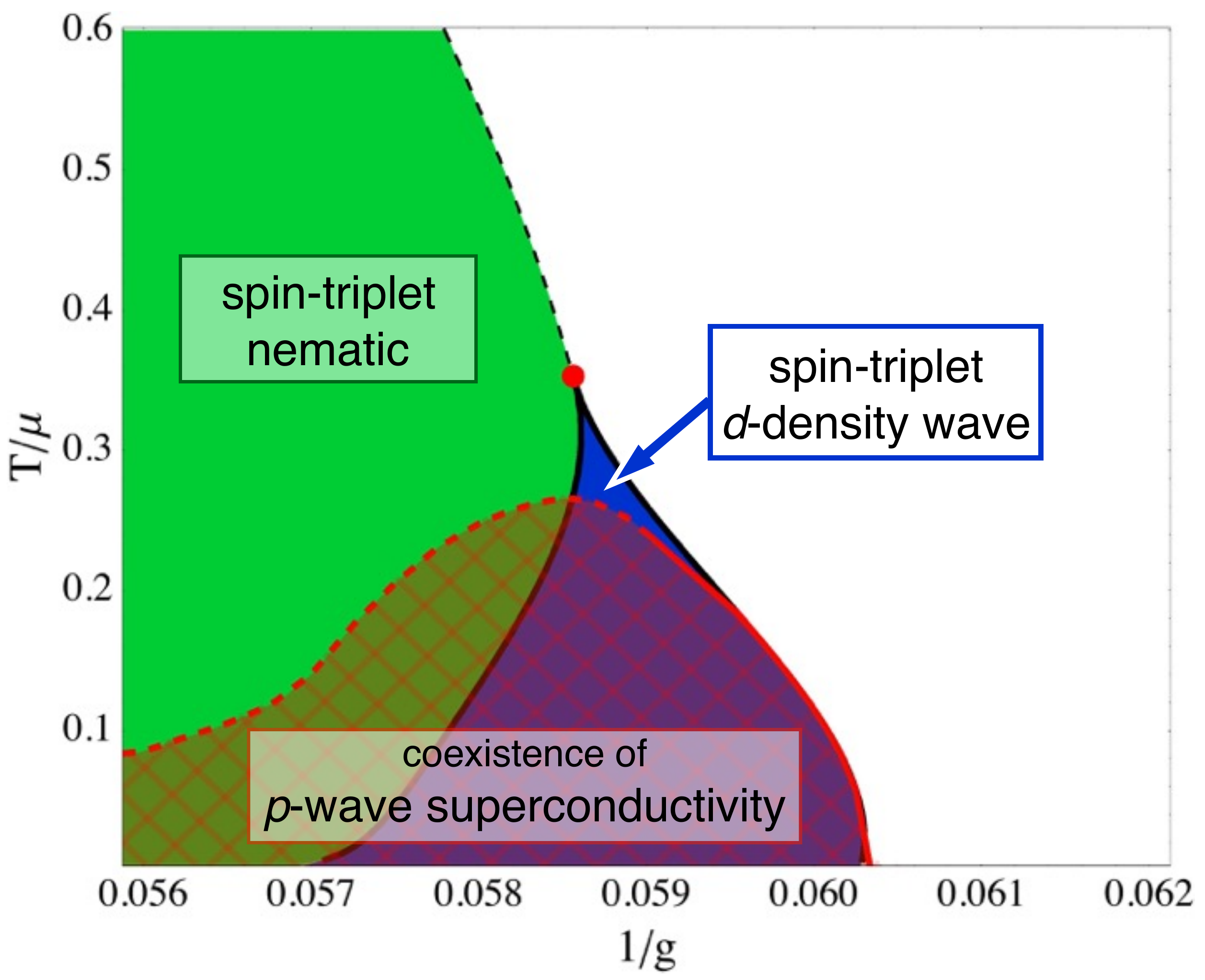}
\caption{Phase diagram as a function of temperature $T/\mu$ and inverse quadrupolar density repulsion $1/g$  in the limit of vanishing interaction range ($\xi=0$). 
At temperatures above the tri-critical point (red), the transition from the isotropic metal to the spin-triplet nematic (green region) is continuous. Below the tri-critical point, fluctuations render the phase 
transition first order and stabilize a region of helical spin-triplet  $d$-density wave order  with ordering wave vector $\bq = q/\sqrt{2} \,(1,\pm1,0)$ 
(blue region). The shaded region indicates $p$-wave superconducting order that forms on the background of spin-triplet nematic  or $d$-density wave order.}
\label{fig.phase}
\end{center}
\end{figure}

A mean field analysis of the model described in Section \ref{sec.model} predicts a continuous, Stoner-like Pomeranchuk transition into a $d$-wave spin-triplet nematic phase with quantum critical 
point at some value of the quadrupole interaction strength $g=V(0)$. Allowing for the effects of fluctuations leads to a much richer phase diagram. Using the quantum order-by-disorder approach 
reveals that the quantum critical point is masked by the formation of a region of triplet $d$-density wave order, as shown in Fig~\ref{fig.phase}. $p$-wave superconductivity 
forms in this background with the orientation of the superconducting order parameter locked to that of the spin-triplet nematic order. In the  modulated phase, this leads to an 
exotically intertwined order. 

The computation of this phase diagram proceeds by first finding the global minimum of the total free energy $F(\eta, q)$ [Eq.~(\ref{eq.FEmfq}) and Eq.~(\ref{eq.FEflucq})] and then evaluating the 
superconducting transition temperature on this background from Eq.~(\ref{eq.FayAppel}), as explained in Sec.~\ref{sec.pSc}. 
First we determine the phase boundaries of spin-triplet nematic order. As shown in Sec.~\ref{sec.fluct}, 
fluctuations give rise to a $\ln T$ contribution to the $\eta^4$ coefficient, rendering the nematic transition discontinuous at low temperatures. The tri-critical point at which the order of the transitions 
changes, is determined by the intersection of the lines along which the coefficients of $\eta^2$  and $\eta^4$ vanish, Eqs.~(\ref{eq.tricrit}). 

For a vanishing range of interactions, $\xi=0$ [see Eq.~(\ref{eq.int})], the tri-critical point is located at  $1/g_c\simeq 0.0586$ and $T_c/\mu\simeq 0.35$. In real materials, disorder \cite{Belitz+99,Ho+2008} 
and the finite range of the interactions~\cite{Keyserlingk+2013} reduce the relative strength of the fluctuation contributions (\ref{eq.FEflucq}), leading to an exponential suppression of the first-order 
behavior. The exponential decrease of the tri-critical temperature as a function of the interaction range $\xi$ follows immediately from Eq.~(\ref{eq.int}) and the asymptotic low-temperature behavior of Eq.~(\ref{eq.tricrit}), yielding $T_c \sim \exp\left\{-\beta_4(T=0)/[2c \llangle \Phi_1^4 \rrangle 
V^2(2 k_F)]\right\}$.

Since the $\eta^4$ and $q^2\eta^2$ coefficients change sign simultaneously, fluctuations stabilize a  triplet $d$-density wave state  below the tri-critical point. The region of the modulated phase is much larger than in the case of an itinerant ferromagnet. This is a consequence of the different behavior of the  non-analyticities as $T\to0$. 

We must also account for different orientations of the helical ordering vector ${\bq}$. Minimizing the free energy for different orientations of $\bq$ along high-symmetry directions relative to the 
deformation $\Phi_1(\bk)=k_x^2-k_y^2$ \cite{qdirec}, we find that for all values of $T$ and $g$ over which fluctuations stabilize modulated order, the helical triplet $d$-density
wave with 
$\bq=q(1,1,0)/\sqrt{2}$ has the lowest free energy.

The transitions between the modulated and homogeneous ordered states is continuous, but would become weakly first-order in the presence of magnetic anisotropy. 
Our theory predicts that the transition between the isotropic metal and the triplet $d$-density wave is discontinuous. Such first-order behavior is expected for phases that are stabilized by the order-by-disorder (or Coleman-Weinberg) mechanism, especially in metals where the fluctuations are not associated with an isolated  point in momentum space but with particle-hole excitations around the entire
 Fermi surface. 

The region of $p$-wave superconducting order is calculated by assuming a continuous transition in the spin-nematic background, using Eq.~(\ref{eq.FayAppel}). Superconducting pairing is strongly enhanced by the spin-triplet nematic or $d$-density wave order and the superconducting dome is therefore almost completely contained within the ordered spin-triplet states (see Fig.~\ref{fig.phase}). 
Note that outside the ordered regions $T_\textrm{SC}$ drops to exponentially small values. This behavior is very similar to the $p$-wave superconductivity forming on the background
of $s$-wave ferromagnetism \cite{Conduit+13} and consistent with experimental observations \cite{Saxena+00}. 

As noted in Section \ref{sec.pSc}, mode-mode coupling locks the orbital $d$-wave form factor and the superconducting $p$-wave order parameter in the relative orientation shown in Fig.~\ref{fig.orientation}. In the region of overlap between superconductivity and triplet $d$-density wave order this causes a spatial modulation of the superconducting order parameter, 
giving rise to a much-sought \emph{pair density wave} state. 

\begin{figure}[t!]
\begin{center}
\includegraphics[width= 0.9 \linewidth]{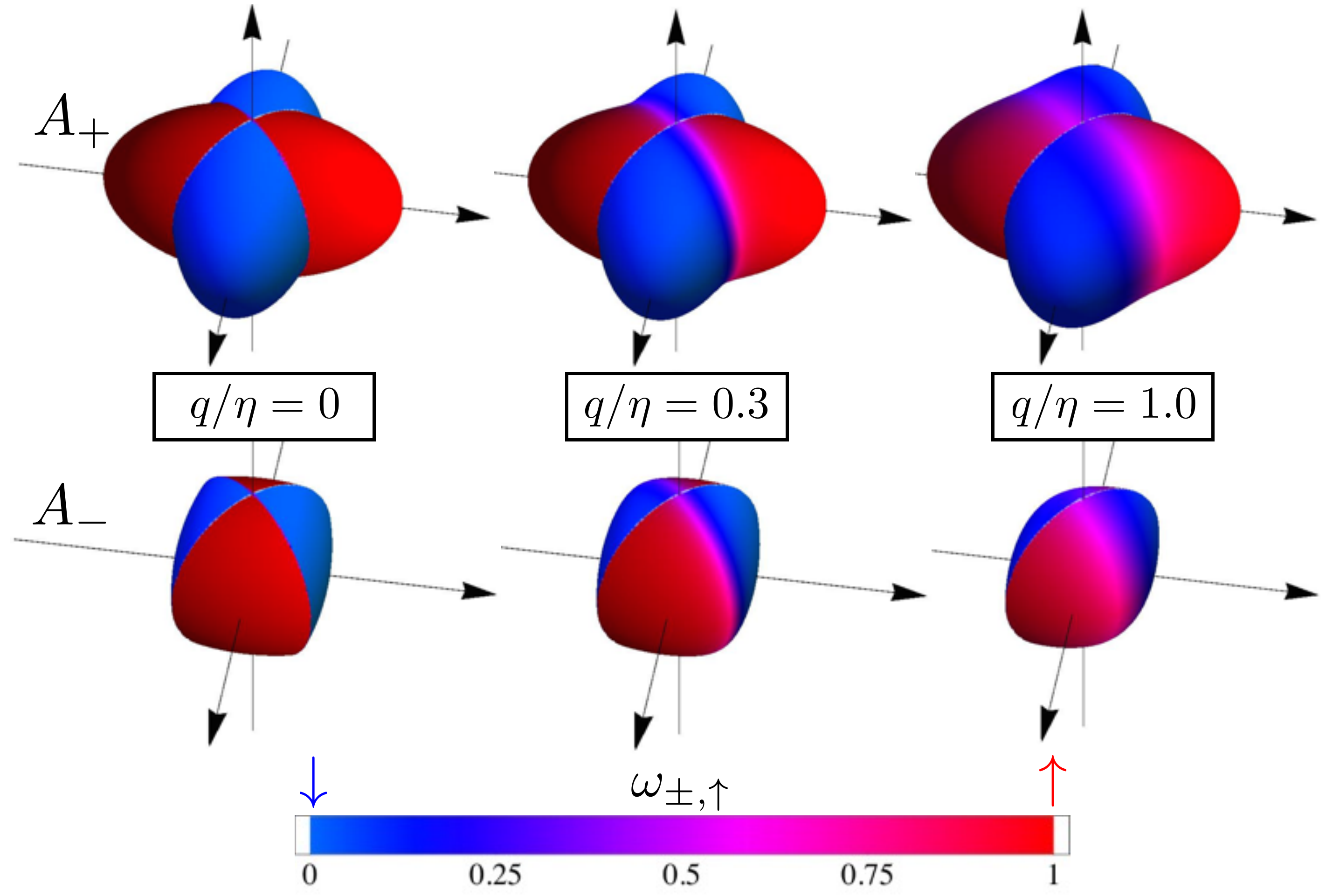}
\caption{Fermi surfaces $A_\pm$ of the electronic bands $\epsilon_\pm(\bk)$  in the presence of helical spin-triplet $d$-density wave order with $\bq=q(1,1,0)/\sqrt{2}$. 
Red and blue colors denote the spin-up and spin-down character of the lobes. As we increase the value of $q$, moving from left to right, we see that this spin character gets mixed, and the 
Fermi surfaces deform along the $(1,1,0)$ direction.}
\label{fig.momentum}
\end{center}
\end{figure}

\subsection{Visualization in momentum- and real-space}
The helical spin-triplet $d$-density wave is not easy to visualize.  For small $\bq$ vectors, corresponding to a long period of the 
modulation in real space, a Wigner representation as used in Fig.~\ref{fig.schematic} is the most convenient depiction. This is a mixed real/momentum space representation. 
Over a subsystem whose size is less then the wavelength of the modulation, the order is approximately uniform and one may define a quasi Fermi surface equivalent to that of the 
related homogeneous spin-triplet nematic.  The helical modulation in spin space implies that the spin direction rotates from sub-system to sub-system with a period $2 \pi /q$. 

A purely momentum space picture is also useful as it helps reveal how spatial modulation might be favored by the softening of fluctuations. In 
Fig.~\ref{fig.momentum}, we show the Fermi surfaces $A_+$ and $A_-$ for the two electronic bands
 $\epsilon_\pm(\bk)=k^2\mp\sqrt{(\bk.\bq)^2+\eta^2  \Phi_1^2(\bk)}$, 
 with the wavevector ${\bq}$ in the favored $(1,1,0)$ direction. $A_+$ and $A_-$ are the Fermi surfaces for electrons with spin parallel and anti-parallel to the background helimagnetic 
 ordering, respectively. In the limit $\bq=0$, we recover the elliptical Fermi surfaces of the homogeneous spin-triplet nematic. These deformations of the Fermi surfaces change the spectrum of 
 electronic particle-hole excitations and enhance the phase space for fluctuations. 

\begin{figure}[t!]
\begin{center}
\includegraphics[width= 0.9\linewidth]{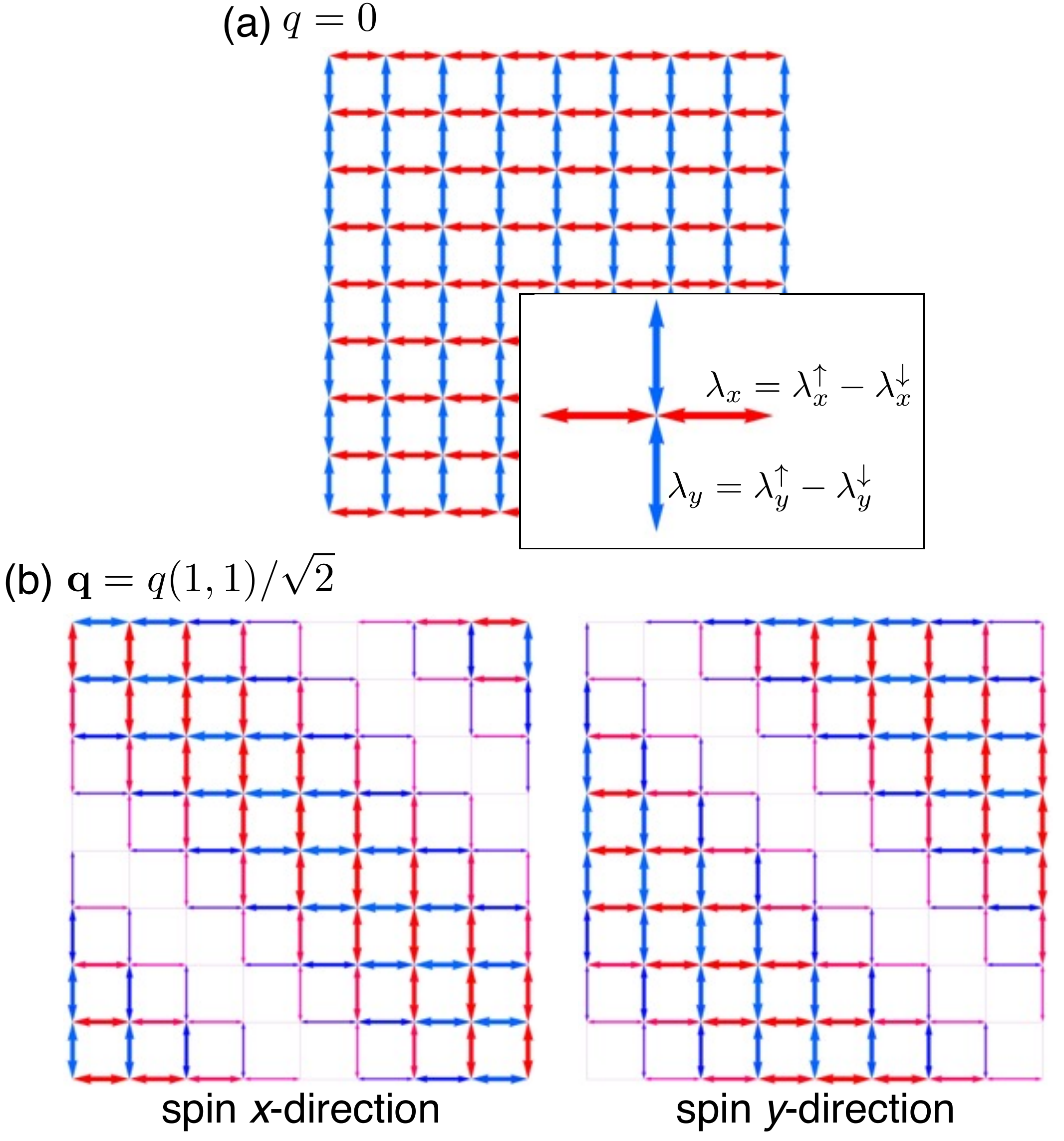}
\caption{Visualization of the spin-triplet nematic order parameter on a square lattice. (a) The homogeneous state corresponds to bond order which
breaks the rotation symmetry of the square lattice. The order parameter changes sign under 90 degree rotation, as well as under spin inversion, and is invariant under 
the two combined operations. (b) Bond-density wave order corresponding to the helical spin-triplet  $d$-density wave with $\bq$ along the $(1,1)$ direction. The two 
panels show the $x$ and $y$ spin components of the modulated order parameter, respectively.}
\label{fig.real}
\end{center}
\end{figure}

We conclude this section by providing a real-space picture of the homogeneous spin-triplet nematic and the modulated triplet $d$-density wave states
when projected onto a lattice. This illustrates the 
connection of our continuum model to lattice based models of bond density wave order. For simplicity, we consider a two-dimensional square lattice. We discretize the order parameter 
$\bm{\eta}(\br)=\langle \hat{\bR}_1^t(\br)\rangle=\frac12 \langle\Psi^\dagger(\br) 
\bm{\sigma} (\partial_x^2-\partial_y^2)\Psi(\br)\rangle$, which (for fixed $\alpha=1$) is a three dimensional vector in spin space. For the homogeneous spin-triplet nematic state along the 
$z$ spin direction, we obtain the lattice order parameter
\begin{equation}
\tilde{\eta} = \left(\lambda_x^\uparrow-\lambda_y^\uparrow\right)-\left(\lambda_x^\downarrow-\lambda_y^\downarrow\right)
\nn
\end{equation}
in terms of expectation values of bond operators, $\lambda_{x(y)}^\nu = \langle \psi^\dagger_{\br,\nu}  \psi_{\br+\hat{\bm{x}}(\hat{\bm{y}}),\nu} \rangle$. The order 
parameter $\tilde{\eta}$ is shown in Fig.~\ref{fig.real}(a). It changes sign under spin inversion, as well as under 90 degree rotation. Because $\langle \hat{R}_1^s(\br)\rangle=0$,
the strain components of spin-up and spin-down electrons exactly cancel each other, $\left(\lambda_x^\uparrow-\lambda_y^\uparrow\right)+
\left(\lambda_x^\downarrow-\lambda_y^\downarrow\right)=0$. 

In the helical spin-triplet  $d$-density wave, the spin direction rotates in a plane in spin space, e.g. between the $x$ and $y$ directions, as specified by the order parameter 
$\bm{\eta}(\br)$ (\ref{eq.helical_order}). This can again be expressed in terms of expectation values of bond-operators, 
\begin{subequations}
\begin{eqnarray}
\tilde{\eta}_x(\br) & = &  \langle\Psi^\dagger_\br \sigma_x  \Psi_{\br+\hat{\bm{x}}}\rangle -  \langle\Psi^\dagger_\br \sigma_x  \Psi_{\br+\hat{\bm{y}}}\rangle=\tilde{\eta} \cos(\bq\br),
\qquad
\nn \\
\tilde{\eta}_y(\br) & = &  \langle\Psi^\dagger_\br \sigma_y  \Psi_{\br+\hat{\bm{x}}}\rangle -  \langle\Psi^\dagger_\br \sigma_y  \Psi_{\br+\hat{\bm{y}}}\rangle=\tilde{\eta} \sin(\bq\br).
\qquad
\nn
\end{eqnarray}
\end{subequations}
The order-parameter components $\tilde{\eta}_x(\br)$ and $\tilde{\eta}_y(\br)$ are shown in Fig.~\ref{fig.real}(b) for a $\bq$ vector along $(1,1)$  that is commensurate with the underlying square lattice. Fig.~\ref{fig.real} is in essence a lattice projection of the Wigner representation shown in Fig.~\ref{fig.schematic}.

\section{Experimental signatures of the spin-triplet nematic}
\label{sec.exp}

Spin-triplet nematic order simultaneously breaks spatial rotation symmetry and spin-rotation symmetry. This entanglement of spin and spatial degrees of freedom has important consequences for measurements. The addition of translational symmetry breaking in the helical spin-triplet $d$-density wave adds further potential for observation. 

 It is important to note that spin-triplet nematics are very different from -- and potentially easier to observe than -- nematics in spin space (often called spin nematics) \cite{spinnematic,Chen+1971,Andreev+1984,Chubukov1991, Shannon+2006}. They are also distinct from charge nematics, which are observable, for example by resistive anisotropy measurements \cite{Grigera12112004}. Both of these other orders are invisible to the probes that we discuss here. 
 
\begin{figure}[t!]
\begin{center}
\includegraphics[width= \linewidth]{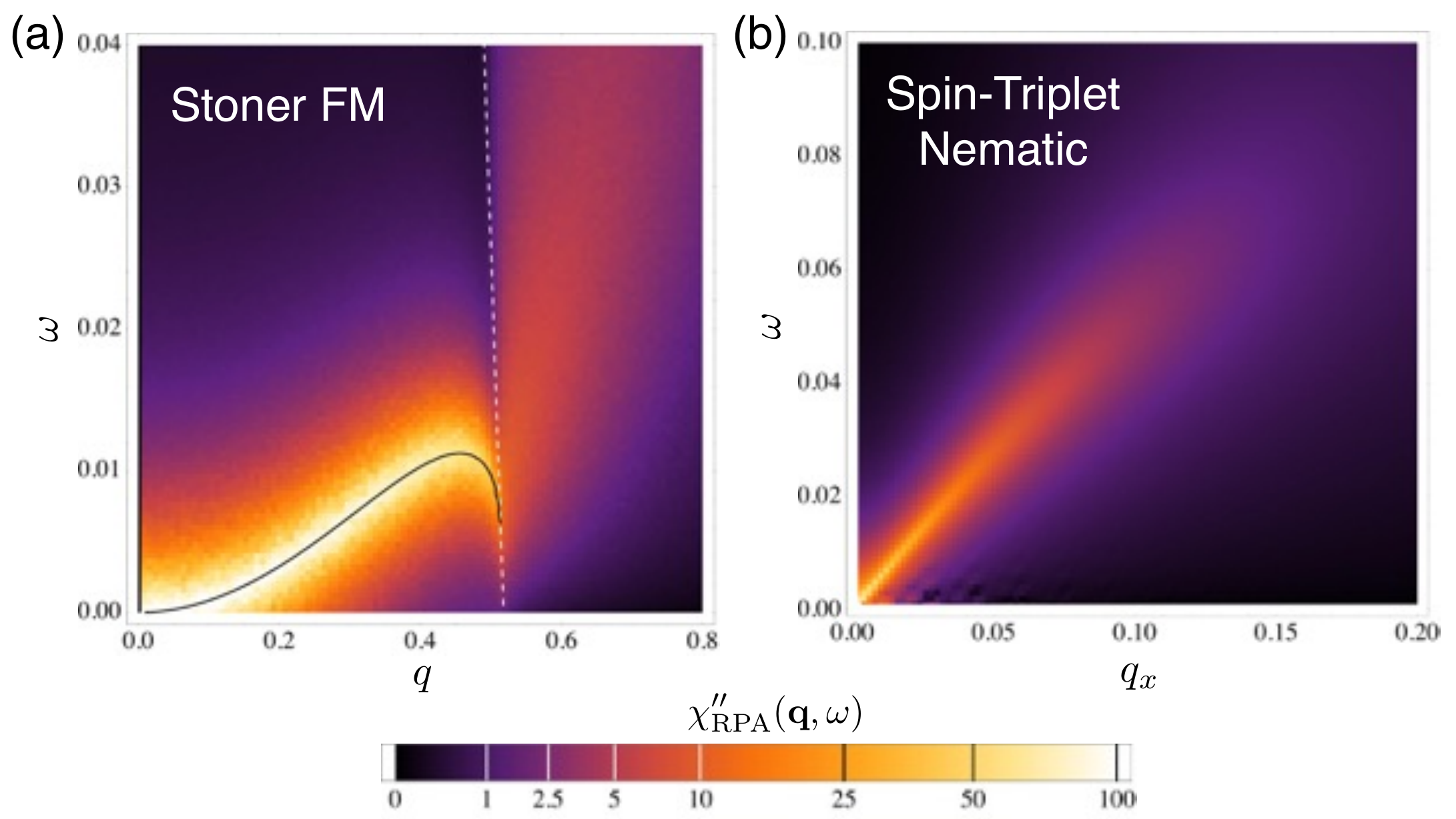}
\caption{Comparison of the magnetic excitation spectra of (a) the Stoner ferromagnet and (b) the spin-triplet nematic. The color gradient shows
the imaginary parts of the magnetic RPA susceptibilities, $\chi''_\textrm{RPA}(\bq,\omega)$, calculated numerically using expressions given in Appendix \ref{appendix_SpinSusceptibility}.  
(a) The ferromagnet exhibits sharp spin-wave excitations with dispersion $\omega\sim q^2$  (thin black line) that become damped as they enter the Stoner continuum at 
$\omega_{ph}=2 U m + q^2 - 2 q \sqrt{1 + U m}$ (dashed white line).  
(b) For the spin-triplet nematic there is no gap below the 
particle-hole continuum and the magnetic excitations are always damped. They follow a linear dispersion relation, $\omega\sim q$. In calculating this figure, we have used a grid 
of $1000^3$ $\bk$-points. We fix $U m/\mu=g\eta/\mu=0.5$ and $T/\mu=0.005$. A physically-insignificant broadening $\delta=0.0005$ was used to improve convergence.}
\label{fig.RPA}
\end{center}
\end{figure}

Let us first study the static response. We assume for simplicity that in the disordered phase 
the system is tetragonal with $x$ and $y$ directions degenerate, and that the nematic distortions are along $x$ and $y$, as shown in Fig.~\ref{fig.dist}(a). While charge (or spin-singlet) nematic 
order breaks the symmetry between $x$ and $y$ directions and induces an orthorhombic distortion, the spin-triplet nematic phase remains tetragonal, since the Fermi-surface deformations for 
the two spin species are of opposite sign. This leads to perfect cancellation of the corresponding strain components. 

The coupling between spin and spatial degrees of freedom can be seen experimentally if either a magnetic field or strain is applied to the system. As pointed out in Ref.~\cite{Wu+2007}, a magnetic field unbalances the two spin species, generating a strain field and resulting in a small orthorhombic lattice distortion. Conversely,  breaking the tetragonal symmetry by applying strain along either $x$ or $y$ changes the ellipticities of 
the Fermi surfaces in opposite ways and induces a small magnetic moment. These responses can be extremely small, however.

A helical modulation of the spin-triplet order can enhance these signatures. When a uniform field is applied in this case, the strain response inherits the spatial modulation which could be 
visible  in high resolution diffraction experiments.  Since the signature is shifted away from other uniform effects that may 
occlude its measurement, it should be more unambiguously observable. 

The study of Larkin-Ovchinikov-Fulde-Ferrell physics has proven difficult in bulk materials because of the small parameter regime over which they exist. This has been circumvented in some situations by using heterostructures to enforce a length scale and proximity effects at the boundaries to induce the order~\cite{RevModPhys.77.935}. We speculate that similar heterostructures might be used to induce the subtle intertwining of  triplet $d$-density wave  and $p$-wave superconducting order that we propose; twisted ferromagnetic capping layers may tip a candidate material into the helical phase, with accompanying signatures in transport signaling $p$-wave superconductivity.

Finally, we note that the dynamical susceptibility of the spin-triplet nematic  has some distinctive features that are potentially observable in experiment (see Fig.~\ref{fig.RPA} for a comparison 
with a metallic ferromagnet). Calculation of the dynamical susceptibility at the level of the RPA approximation (see Appendix~\ref{appendix_SpinSusceptibility}) shows the expected linear 
dispersion of excitations, but with a surprising non-linear, non-Landau damping, $\Gamma(q) \sim q^2$, in contrast to the conventional, linear, Landau damping, $\Gamma(q)\sim q$ of the 
ferromagnet. This signature is potentially observable in neutron scattering, especially when shifted to finite wave-vector due to a helical modulation.

\section{Discussion}
\label{sec.disc}
Spin-triplet nematic order has a number of interesting static and dynamical properties. In the $d$-wave channel, it is characterized by elliptical distortions of the Fermi surface that have 
opposite sign for different spin components. This static order induces a cross response between magnetic and stain channels; an applied magnetic field unbalances the spins and leads 
to a net orthorhombic distortion. The fluctuations about the spin-triplet nematic state have a linear dispersion, unusual non-Landau damping and characteristic quantum critical properties. 
Since they couple to spin, they have the potential to be seen in neutron scattering experiments. 

The fluctuations may also drive new physics that has not been studied to date. We have shown how fluctuations can self-consistently stabilize a phase of spin-triplet $d$-density wave order with a helical modulation of spin. A uniform magnetic field applied to this modulated state  can in principle drive a spatially modulated strain response - a response both in a different channel and at a different wave-vector. Moreover, this behavior can be further intertwined with $p$-wave superconducting order.

The fluctuation-driven formation of $d$-density wave order stems from the same fermionic quantum order-by-disorder mechanism that is responsible for the formation of spiral magnetic order in itinerant ferromagnets~\cite{Conduit+09,Karahasanovic+12,Kruger+12,Thomson+13,Pedder+13,AbdulJabbar+2015}. We have demonstrated these features using an idealized single-band model of electrons that interact through a quadrupole density-density repulsion. A mean field analysis of this model predicts a Pomeranchuk instability to $d$-wave spin-triplet nematic order~\cite{Wu+2007} akin to the Stoner transition of the itinerant ferromagnet. The similarities persist when analyzing the effects of fluctuations; terms in the Ginzburg-Landau expansion of the free energy of spin-triplet nematic order are related to those of the ferromagnet supplemented with appropriate angular averages of orbital form factors. The additional angular dependencies lead to a modification of the non-analyticities of the Ginzburg-Landau expansion  of the spin-triplet nematic compared to that of the ferromagnet. The region of parameter space occupied by fluctuation-induced behavior is larger for the spin-triplet nematic than for its ferromagnetic analogue. 

The triplet $d$-density wave is essentially a continuum version of bond density wave order~\cite{Metlitski+10a,Metlitski+10b,Efetov+13,Wang+14}. A number of 
analyses of the latter have studied models in which band structure plays 
an intimate role, enhancing susceptibility to order at finite wave vector \cite{Khavkine+2004,Kee+2005,Yamase+2005,Yamase2013}. This is  reminiscent of the case of ferromagnetic order, which 
can also be driven helimagnetic by density-of-states effects \cite{Berridge+2010}. We have demonstrated  a new way to achieve complex, spatially modulated order, that does not require such 
features in the density of states, breaking of inversion symmetry, or frustration. This mechanism is rather independent of microscopic details, e.g. tight-binding corrections to the dispersion 
do not qualitatively change the phase diagram, 
as long as the system is far from instabilities due to nesting or van Hove singularities \cite{Karahasanovic+12}. 

Electronic models that contain interactions only in  a single higher angular momentum  ($l \neq 0$) channel are 
highly idealized. They are designed to exhibit electronic nematic phases and to study the instabilities of such phases. In real materials the electron-electron interactions are composed of different angular momentum channels. Indeed, in systems where the effective interaction is peaked at momenta near to $2k_F$, signaling a tendency towards local crystallinity,  it may 
 be possible for several of the Landau parameters to be large and negative \cite{Oganesyan+2001}. This leads to the intriguing possibility of phase competition or cooperation between instabilities in several different angular momentum channels. An interesting scenario would be the stabilization of a $d$-wave spin-triplet nematic in a ferromagnetic  background. This would result in a true electron nematic state that couples to strain. Even if the bare higher angular momentum components of the interaction are negligible, fluctuations can dynamically generate effective interactions and resulting instabilities in higher-angular momentum channels \cite{Karahasanovic+12,Chubukov+09}. 

The fermionic quantum order-by-disorder approach allows us to address the stability of new and exotic phases of matter by focussing upon the effect of order upon the fluctuation spectrum. By expressing the free energy of the system as a functional of the electron dispersion in the presence of various broken-symmetry states, this method allows us to study the competition and cooperation between several phases. The appearance of exotic orders like the helical spin-triplet $d$-density wave  from simple models without  Fermi-surface nesting and frustration emphasizes the important role that quantum fluctuations may play in the low temperature properties of interesting materials.

{\bf{Acknowledgements}}
This work has been supported by the EPSRC through grant 
EP/I004831/2.  CJP is supported by the National Research Fund, Luxembourg under grant ATTRACT 7556175.

\appendix

\section{Angular averages}
\label{appendix_angular}
Here we give explicit expressions for the various angular averages that enter in the coefficients of the free energy. In three dimensions, the angular average of a function $\Phi(\phi,\theta)$ of spherical angles is defined as
\begin{equation}
\llangle \Phi(\phi,\theta) \rrangle = \frac{1}{4\pi}\int_0^{2\pi}\ud\phi \int_0^\pi \ud\theta \sin\theta \; \Phi(\phi,\theta).
\nn
\end{equation}
For the homogeneous spin-triplet nematic state we have to compute angular averages of powers of $\Phi_1(\hat{\bk})=\hat{k}_x^2-\hat{k}_y^2=\cos(2\phi)\sin^2\theta$.
Such averages factorize into elementary integrals,
\begin{eqnarray}
\lefteqn{\llangle \Phi_1^m(\hat{\bk})\rrangle 
=
 u_m v_m,}
& &
 \nn
 \\
u_m & = & 
\int_0^{2\pi}\frac{\ud \phi}{2\pi}   \cos^m (2\phi)   
= 
\left\{\begin{array}{cc} \frac{(m-1)!!}{(m)!!} & m\phantom{.}\textrm{even} \\ 0 & m\phantom{.}\textrm{odd}
 \end{array}\right.  
 \nn
 \\
v_m 
& = &  
\frac 12 \int_0^{\pi}\ud\theta \sin\theta  (\sin^2\theta)^m     
=\frac{\sqrt{\pi}}{2} \frac{\Gamma(m+1)}{\Gamma(m+\frac32)}.
\nn
\end{eqnarray}
Here $\Gamma(x)$ denotes the Gamma function and the double factorials are defined as $(2m)!!=(2m)\cdot(2m-2)\cdot\ldots\cdot4\cdot2$
for even numbers and $(2m-1)!!=(2m-1)\cdot(2m-3)\cdot\ldots\cdot3\cdot1$ for odd numbers, respectively. 

Allowing for spatial modulation of the spin-triplet nematic order, we must also calculate the averages
$ \llangle  \Phi_1^m(\hat{\bk}) (\hat{{\bk}}.{\hat{\bq}})^n  \rrangle$
for $n=2$ and $n=4$. 

\noindent
For $\bq=q(0,0,1)$, we obtain 
\begin{eqnarray*}
\llangle    \Phi_1^{2m}(\hat{\bk}) (\hat{{\bk}}.{\hat{\bq}})^2\rrangle 
& = & 
u_{2m} \left(v_{2m}-v_{2m+1}\right), 
\\ 
\llangle   \Phi_1^{2m}(\hat{\bk}) (\hat{{\bk}}.{\hat{\bq}})^4\rrangle 
& = & 
u_{2m}  \left(v_{2m}-2v_{2m+1}+v_{2m+2}\right),
\end{eqnarray*}

\noindent
For $\bq=q(1,1,0)/\sqrt{2}$, we obtain 
\begin{eqnarray*}
\llangle    \Phi_1^{2m}(\hat{\bk})(\hat{{\bk}}.{\hat{\bq}})^2\rrangle 
 & = & 
 \frac 12 u_{2m} v_{2m+1},\\
\llangle   \Phi_1^{2m}(\hat{\bk})  (\hat{{\bk}}.{\hat{\bq}})^4\rrangle  
& = &  
\frac 14 \left(2 u_{2m}-u_{2(m+1)}\right) v_{2m+2},
\end{eqnarray*}

\noindent
and finally, for $\bq=q(1,0,0)$, we obtain
\begin{eqnarray*}
\llangle  \Phi_1^{2m}(\hat{\bk})   (\hat{{\bk}}.{\hat{\bq}})^2\rrangle 
& = & 
\frac 12 u_{2m} v_{2m+1}, \\
\llangle   \Phi_1^{2m}(\hat{\bk})  (\hat{{\bk}}.{\hat{\bq}})^4\rrangle  
 & = & 
 \frac 14 \left(u_{2m}+u_{2(m+1)}\right) v_{2m+2}. 
\end{eqnarray*}

\section{Spin Susceptibility in the Presence of Ferromagnetic and Spin-triplet Nematic Order}
\label{appendix_SpinSusceptibility}
All of the novel features of the spin-triplet nematic are driven by the nature of the spin fluctuations that it supports. 
Bare expressions for the key quantities that we need are given by 
\begin{eqnarray}
\chi^{\uparrow \downarrow}_0(\bq,\omega) 
&=&
 \sum_\bk \frac{n_F[\epsilon_\uparrow(\bk)]-n_F[\epsilon_\downarrow(\bk+\bq)]}
{\epsilon_\downarrow(\bk+\bq)-\epsilon_\uparrow (\bk)-(\omega+i 0_+)}
\\
\chi^{\downarrow \downarrow}_0(\bq,\omega) 
&=&
 \sum_\bk \frac{n_F[\epsilon_\downarrow(\bk)]-n_F[\epsilon_\downarrow(\bk+\bq)]}
{\epsilon_\downarrow(\bk+\bq)-\epsilon_\downarrow (\bk)-(\omega+i 0_+)}.
\label{eq.chi0}
\end{eqnarray}
This expression takes the same form for both the spin-triplet nematic and the itinerant ferromagnet, the distinction between the two arising from the different mean field electron dispersions. 
For the  spin-triplet nematic, the electron dispersion is given by $\epsilon_\nu(\bk)=k^2-\nu  g f(\bk)\Phi_1(\bk) \eta$. For the Stoner ferromagnet, $\epsilon_\nu(\bk)=k^2-\nu U m$ with $U$ the 
conventional Coulomb density-density repulsion and $m$ the magnetization. In the latter case, it is possible to calculate the bare susceptibilities analytically with the results \cite{Moriya85}

\begin{figure}[t]
\begin{center}
\includegraphics[width= \linewidth]{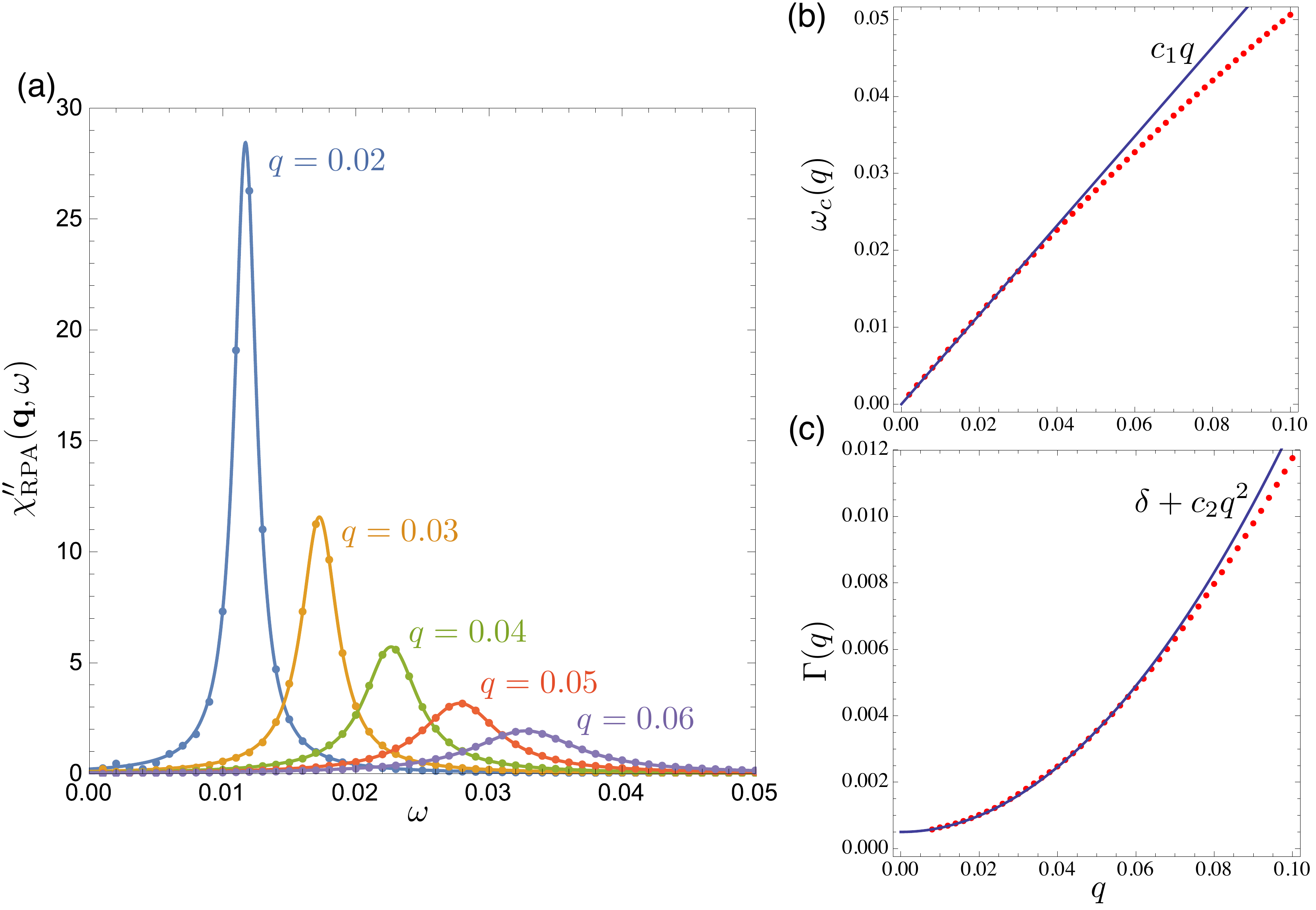}
\caption{Dispersion and Damping of Magnetic Fluctuations in the Spin-Triplet Nematic. (a) Magnetic susceptibilities $\chi''_\textrm{RPA}(\omega)$ of the spin-triplet nematic for a few fixed values of $q=q_x$. Solid lines show 
Lorentzian fits.
(b) Dispersion relation $\omega_c(q)$ extracted from the maxima of Lorentzian fits. For small propagation 
vectors, the dispersion is linear (blue line). 
(c) Damping rate $\Gamma(q)$ extracted from the 
Lorentzian fits. We find
non-Landau damping, $\Gamma(q)=c_2 q^2+\delta$ (blue curve),  where $\delta=0.0005$ is the physically insignificant broadening we introduced to improve convergence of 
the numerical integration.}
\label{fig.damping}
\end{center}
\end{figure}

\begin{eqnarray}
{\cal R}e \; \chi^{\uparrow \downarrow}_{0,\textrm{FM}}({\bq}, \omega) & = & \sum_{\nu=\pm} \nu \left\{\frac{4q^2\mu_\nu-\left(\nu q^2+2Um-\omega\right)^2}{64 \pi^2 q^3}\right.\nn\\
& & \times \ln\left| \frac{\nu q^2+2Um -\omega +2q\sqrt{\mu_\nu}}{\nu q^2+2Um -\omega -2q\sqrt{\mu_\nu}}  \right|\nn\\
& & \left.+ \frac{(\nu q^2+2Um-\omega)\sqrt{\mu_\nu}}{16 \pi^2 q^2}\right\}\\
{\cal R}e \; \chi^{\downarrow \downarrow}_{0,\textrm{FM}}({\bq}, \omega) & = & \sum_{\nu=\pm} \frac{4q^2\mu_\downarrow-\left(q^2-\nu \omega\right)^2}{64 \pi^2 q^3}\nn\\
& & \times \ln\left| \frac{ q^2-\nu\omega+2q\sqrt{\mu_\downarrow}}{ q^2-\nu\omega-2q\sqrt{\mu_\downarrow}}  \right| + \frac{\sqrt{\mu_\downarrow}}{8 \pi^2}
\label{Eq.chi02}
\end{eqnarray}
where $\mu_\nu=\mu+\nu Um$. The orbital factors entering {\it via} the mean field dispersion lead to qualitative differences in the bare susceptibilities of the spin-triplet nematic. 
They also render the integrals much more difficult. Some progress can be made in calculating $\chi^{\downarrow\downarrow}({\bf q},0)$. 
 If we make the approximation $f(\bk)\Phi_1(\bk)\simeq \Phi_1(\bk)$,  the integral can be carried 
out by rescaling $\tilde{k}_x=k_x\sqrt{1-\nu \eta}$ and $ \tilde{k}_y=k_y\sqrt{1+ \nu \eta}$ 
so that $\epsilon_\nu=|\tilde{\bk} |^2$. After this we obtain ${\cal R}e\chi^{\downarrow \downarrow}_0 (\bq,0)= {\cal R}e\chi^{\downarrow \downarrow}_{0,\textrm{FM}}(\tilde{\bq},0)/\sqrt{1-\eta^2}$, 
where $\tilde{\bq}$ is 
a suitably rescaled momentum. Care must be taken with this expression. It is only valid at small $\eta$, since it harbors an unphysical divergence of electron density as $\eta \rightarrow 1$. 
$\chi^{\uparrow \downarrow}$ is even trickier and, whilst some analytical progress can be made using similar manipulations, ultimately we resort to numerical evaluation of the integrals. 

\begin{figure}[t]
\begin{center}
\includegraphics[width= \linewidth]{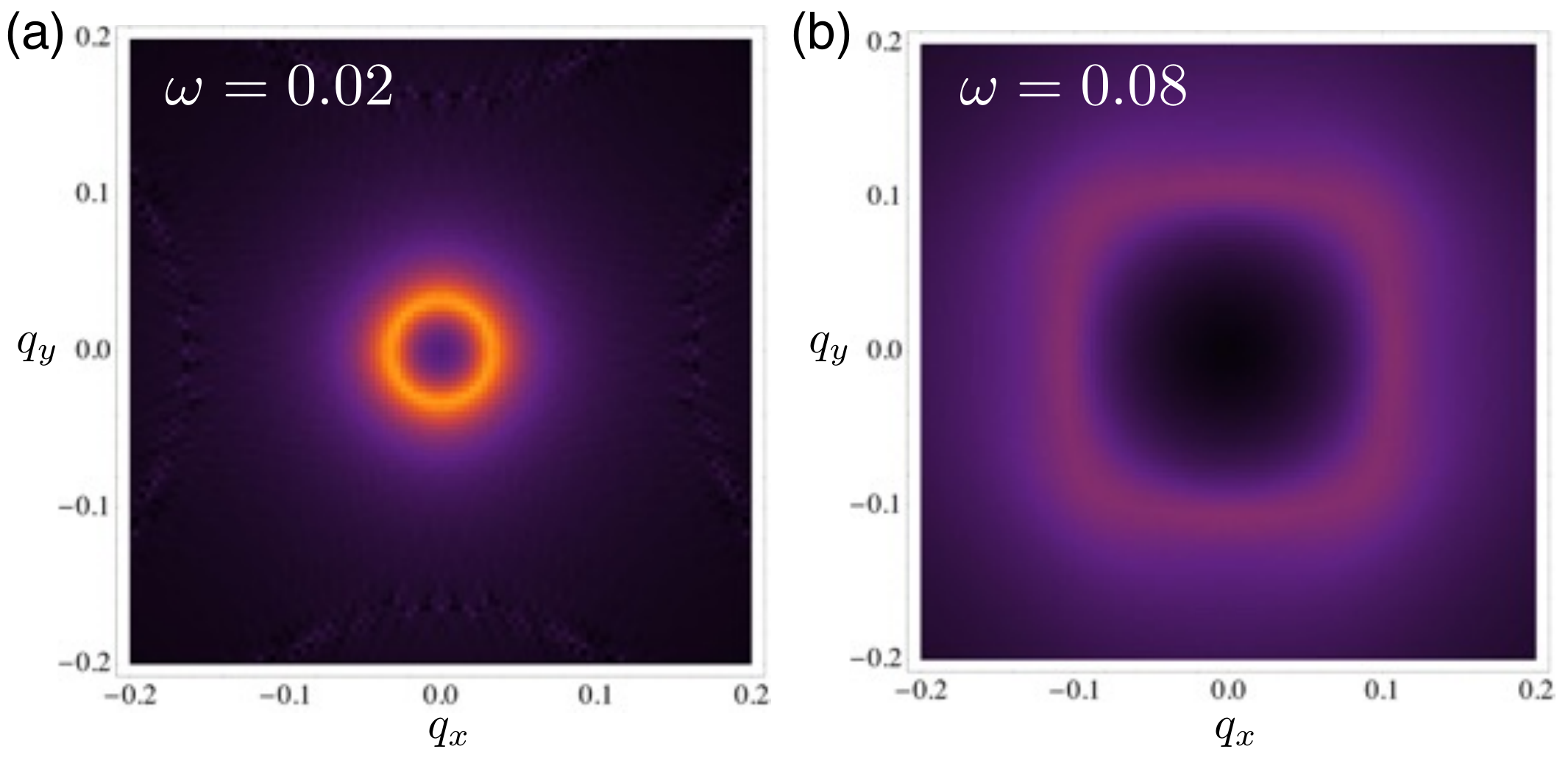}
\caption{Direction Dependence of Magnetic Dispersion in the Spin-Triplet Nematic. Constant energy cuts through the magnetic excitation spectrum of the spin-triplet nematic. 
(a) At low energy ($\omega=0.02$) the excitations are nearly isotropic, forming a well defined ring-like structure in the $q_x$-$q_y$ plane. (b) At higher energy ($\omega=0.08$), 
the excitations have a four-fold, square-like symmetry. Moreover, the intensity is significantly reduced and the excitations are much broader.}
\label{fig.directional}
\end{center}
\end{figure}

Going beyond the bare susceptibility reveals further differences between the ferromagnet and spin-triplet nematic. The RPA susceptibility allows us to determine the dispersion and damping of magnetic fluctuations, which may potentially be probed directly by neutron scattering. For the ferromagnet, the RPA susceptibility is given by the familiar expression
$\chi^{\uparrow \downarrow}_\textrm{RPA}(\bq,\omega) =\chi^{\uparrow \downarrow}_0(\bq,\omega)/[1-U \chi^{\uparrow \downarrow}_0(\bq,\omega)]$. 
For the spin-triplet nematic, the quadrupolar density-density interaction driving the 
instability modifies the RPA expression. This takes the form, 
$\chi^{\uparrow \downarrow}_\textrm{RPA}(\bq,\omega) =\chi^{\uparrow \downarrow}_0(\bq,\omega)/[1-g \tilde \chi^{\uparrow \downarrow}_0(\bq,\omega)]$,
where $\tilde \chi^{\uparrow \downarrow}_0(\bq,\omega)$ is defined as $\chi^{\uparrow \downarrow}_0(\bq,\omega)$ in Eq.~(\ref{eq.chi0}), but with an additional factor of 
$\Phi_1(\bk) \Phi_1(\bk+\bq)$ in the integrand. This additional factor and the different electron dispersion are responsible for a different dispersion and damping rate of the magnetic 
excitations~\cite{Wu+2007}.

The dispersions for the ferromagnet and spin-triplet nematic (given by the resonance conditions  $U {\cal R}e\chi_0(\bq,\omega)=1$ and $g {\cal R}e\tilde \chi_0(\bq,\omega)=1$) are quadratic 
and linear in $q$, respectively. The ferromagnet exhibits conventional Landau damping \cite{Halperin+1969,Chubukov+2014}, $\Gamma(\bq) \sim |\bq|$, whereas the spin-triplet nematic 
displays an unusual non-Landau, non-linear damping $\Gamma(\bq) \sim |\bq|^2$. These results are illustrated in Fig.~\ref{fig.RPA} and Fig.~\ref{fig.damping}, which compare numerical evaluation 
of the dynamical susceptibility of the spin-triplet nematic with that obtained analytically for the ferromagnet. 
Since the spin-triplet nematic order breaks the spatial rotation symmetry, the excitations are expected to be anisotropic. In Fig.~\ref{fig.directional}, constant energy cuts in the $q_x$-$q_y$ plane 
are shown. While for small energies, the excitations are nearly isotropic, with well defined ring-like structures in momentum space, at higher energies, a significant four-fold anisotropy develops.

\section{Superconducting Pairing Function and Field Renormalization}
\label{appendix_Scangular}

\noindent{\it a) Pairing function}: In order to calculate the pairing function, we must perform an appropriate average of the susceptibility over the Fermi surface of the pairing electrons. In the ferromagnet, 
this is given by
\begin{eqnarray}
\lefteqn{
\langle \langle
\Theta_{{\bk}+{\bq}} 
\Theta_{\bk}
 {\cal R}e \chi^{\downarrow \downarrow} ({\bq}, \epsilon_{{\bk}+{\bq}}-\epsilon_{{\bk}})
\rangle \rangle}
& &
\nn\\
&=&
\sum_{{\bk},{\bq}}
\Theta_{{\bk}+{\bq}} 
\Theta_{\bk}
 {\cal R}e \chi^{\downarrow \downarrow} ({\bq}, \epsilon_{{\bk}+{\bq}}-\epsilon_{{\bk}})\nn\\
 & & \times
 \delta(\epsilon_{\bk}^\uparrow -\mu)
 \delta(\epsilon_{{\bk}+{\bq}}^\uparrow -\mu),
 \label{PairingFunctionAppendix}
 \end{eqnarray}
and the angular dependence comes entirely from the $p$-wave factors $\Theta_\bk$ of the superconducting order parameter. The angular integrals can be carried out analytically, leading to the  
 result of Fay and Appel~\cite{Fay:1980kx}. In a spin-triplet nematic background, the spin-susceptibility acquires an angular dependence and the pairing function is modified.
 
 The delta functions restrict the pairing function to its zero frequency part, $\epsilon_{{\bk}+{\bq}}^\uparrow-\epsilon_{{\bk}}^\uparrow = 0$.  A complementary approximation 
 scheme \cite{Roussev2001} neglects the momentum dependence and instead analyses the full frequency dependence. Substituting Eq.~(\ref{eq.chi0}) into (\ref{PairingFunctionAppendix}) and 
specializing to the electron dispersion in the presence of spin-triplet nematic order yields a fairly tricky integral. Luckily, at small $\eta$ we can use the same trick as employed in Appendix~\ref{appendix_SpinSusceptibility}. Approximating the mean-field dispersion by $\epsilon_\nu(\bk)\approx k^2 + \nu  \eta \Phi({\bf k})$ (factor of $g$ absorbed in the redefinition of $\eta$) 
 and rescaling $x$- and $y$-components of momenta as before, permits the radial parts of the momentum integrals to be evaluated. The result is
\begin{widetext}
\begin{eqnarray}
\langle\langle \mathcal{R}e\chi^{\downarrow\downarrow}\rangle\rangle
&=&
 - \frac  {4}{\sqrt{1- \eta^2}^3}
 \sum_{ \Omega_\bp,  \Omega_\bk}
\tilde \Theta_{ \bp}
\tilde \Theta_{ \bk} 
\chi_0\left(\sqrt{\frac{1+\eta}{1-\eta} {(\hat k_x-\hat p_x)^2}+ \frac{1-\eta}{1+\eta}{(\hat k_y-\hat p_y)^2}+  {(\hat k_z- \hat p_z)^2}}\right),\nn
\end{eqnarray}
\end{widetext}
where $\tilde \Theta_\bk$ and  $\tilde \Theta_\bp$ are the transformed $p$-wave form factors after the elliptical rescaling of momenta, $\tilde \Theta_\bk=\Theta_{\tilde \bk}$. $\chi_0(q)$ denotes the susceptibility in the 
absence of any order and at zero frequency (can be obtained from Eq.~(\ref{Eq.chi02}) for $m=0$).

Finally, we expand in powers of the nematic order parameter. The resulting expansion coefficients are messy. Even though they may be calculated analytically, the result is no more revealing than numerical integration over the remaining angular components of momentum. It turns out that the $\Delta^2 \eta^2$ term harbors a zero temperature singularity due the logarithmic divergence of the pairing function at twice the  Fermi momentum. Treating the delta functions as derivatives of Fermi functions cuts off this divergence by shifting the peaks from $\mu$ to $\mu-T$. Allowing for this, the pairing function is given by
\begin{eqnarray}
\langle\langle \mathcal{R}e\chi^{\downarrow\downarrow}\rangle\rangle
=
\lambda\left[ 0.026{+}0.084 \eta+ (0.057{-}0.113 \ln T)\eta^2 \right].
\label{eqn: NSCpf}
\end{eqnarray}
with $\lambda=-\frac{16}{9}/(2 \pi)^6$. The sign and size of the term linear in $\eta$ is a function of the relative orientation of nematic and superconducting symmetry factors. The term 
quadratic in $\eta$ is independent of the relative orientation. Thus, the term linear in $\eta$ determines the preferred alignment of nematic and superconducting order. The result (\ref{eqn: NSCpf})
is for the most favored relative orientation. For $d_{x^2-y^2}$ nematic order, the $p$-wave superconductivity aligns along the $x$-direction (see Fig.~\ref{fig.orientation}). 
Similarly, $p_y$ superconductivity is the most disfavored orientation. 

\vspace{0.1in}
\noindent
{\it b) Field Renormalization}: The field renormalization is given by
\begin{eqnarray}
 \partial_{\epsilon_{\bk}}\Sigma^\uparrow({\bk}, \epsilon_{\bk}) & = &  g^2 \partial_{\epsilon_{\bk}} \sum_{{\bp},{\bq}}
 \frac{n^\downarrow_{{\bp}-{\bq}}  n^\uparrow_{{\bk}-{\bq}} }{ \epsilon^\uparrow_{{\bk}} {+}\epsilon^\downarrow_{{\bp}-{\bq}} {-}\epsilon^\uparrow_{{\bk}-{\bq}} {-}\epsilon^\downarrow_{{\bp}}} \nn\\
 & & +g^2
 \partial_{\epsilon_{\bk}}
 \sum_{{\bp},{\bq}}
 \frac{
  n^\downarrow_{{\bp}} 
\left( n^\downarrow_{{\bp}-{\bq}}  - n^\uparrow_{{\bk}-{\bq}} \right)
}{
 \epsilon^\uparrow_{{\bk}}
 {+}\epsilon^\downarrow_{{\bp}-{\bq}}
 {-}\epsilon^\uparrow_{{\bk}-{\bq}}
 {-}\epsilon^\downarrow_{{\bp}}},\quad
 \label{eq.fieldren}
 \end{eqnarray}
a form that is modified slightly to account for the one-loop regularization of the interaction and split into two parts for calculational convenience. 
For the ferromagnet, the self-energy may be calculated analytically at zero temperature for the ferromagnet. In the case of a background spin-triplet nematic order, the spin susceptibility -
 and hence the self-energy - acquires a direction dependence that has important consequences. 

We approximate the self-energy in the same spirit as our assumption that superconducting pairing is confined to the vicinity of the Fermi surface. This helps us to allow for the directional 
dependence induced by a background spin-triplet nematic order. The approximation amounts to calculating its on-shell value whilst assuming that internal integrals can be linearized at 
the Fermi surface. The latter corresponds to fixing one of the internal legs at the Fermi surface also. Explicitly, we calculate 
$\int d \epsilon_{\bk}  \partial_{\epsilon_{\bk}}
\Sigma^\uparrow ({\bk}, \epsilon_{\bk}) \delta (\epsilon_{\bk}^\uparrow - \mu)$.

We first change the differentiation in the two terms of Eq.~(\ref{eq.fieldren})  to $\epsilon^\uparrow_{{\bk}-{\bq}}$ and $\epsilon^\downarrow_{\bp}$, respectively, 
integrate by parts and linearize at the Fermi surface to obtain
\begin{eqnarray*}
     \partial_{\epsilon_{\bk}}
\Sigma^\uparrow({\bk}, \epsilon_{\bk})
&=&
 -
 \frac  {g^2} 2
 \sum_{ {\bp}}
\partial_\epsilon n^\uparrow_{ {\bp}} \;
 \chi^{\downarrow \downarrow}({\bk}{-}{\bp}, \epsilon^\downarrow_{{\bk}}{-}\epsilon^\uparrow_{{\bp}})
 \nn\\
 & &
{-}g^2 
 \sum_{{\bp}}
  \partial_{\epsilon} n^\downarrow_{{\bp}} \;
  \chi^{\uparrow \downarrow}({\bk}{-}{\bp}, \epsilon^\uparrow_{{\bk}}{-}\epsilon^\downarrow_{{\bp}}).
   \end{eqnarray*}
Treating the derivatives of the Fermi functions as delta functions at the Fermi level and averaging the on-shell value of this field renormalization over the Fermi surface, we obtain
 \begin{eqnarray*}
 \partial_{\epsilon_{\bk}}
\Sigma^\uparrow({\bk}, \epsilon_{\bk})
 \approx
  \frac{1}{2}
  g^2
 \sum_{{\bp},{\bk}}
 \chi^{\downarrow \downarrow}({\bk}- {\bp}, 0)
 \delta( \epsilon_{\bk}^\uparrow- \mu)
    \delta( \epsilon_{{\bp}}^\uparrow- \mu)
\nn
\\
+
 g^2
 \sum_{{\bp},{\bk}}
  \chi^{\uparrow \downarrow}({\bk}-{\bp}, 0)
  \delta( \epsilon_{\bk}^\uparrow- \mu)
    \delta( \epsilon_{\bp}^\downarrow- \mu)
 \end{eqnarray*}
After bringing the field renormalization to this form, we can then compute its contribution to the superconducting transition temperature as
\begin{eqnarray}
\lefteqn{
\langle \langle 
\Theta_{\bk}^2 
\partial_{\epsilon_{\bk}} {\cal R}e \Sigma^\uparrow ({\bk}, \epsilon_{\bk} ) 
\rangle \rangle
}
& &
\nn
\\
&=&
 \frac{1}{2}
  g^2
 \sum_{{\bp},{\bk}}
 \Theta_{\bk}^2 
 {\cal R}e \chi^{\downarrow \downarrow}({\bk}- {\bp}, 0)
 \delta( \epsilon_{\bk}^\uparrow- \mu)
    \delta( \epsilon_{{\bp}}^\uparrow- \mu)
\nn\\
& &
+
g^2
 \sum_{{\bp},{\bk}}
 \Theta_{\bk}^2 
 {\cal R}e \chi^{\uparrow \downarrow}({\bk}-{\bp}, 0)
  \delta( \epsilon_{\bk}^\uparrow- \mu)
    \delta( \epsilon_{\bp}^\downarrow- \mu).\quad
\label{FieldRenormContribution}
\end{eqnarray}

The first term in Eq. (\ref{FieldRenormContribution}) can be analyzed in exactly the same way as the pairing function. As for the pairing function, the term linear in $\eta$  
depends upon the relative orientation of nematic and superconducting order. Additionally, the quadratic $\eta$ term harbors the same $\ln T$ singularity.
The second term in Eq. (\ref{FieldRenormContribution}) requires a bit more work. Unlike $\chi^{\downarrow\downarrow}$, $\chi^{\uparrow \downarrow}$ cannot be evaluated using the approximation $f(\bk)\Phi(\bk) \rightarrow \Phi(\bk)$.  Instead we expand $\chi^{\uparrow \downarrow}$ explicitly to quadratic order in $\eta$,
\begin{eqnarray}
\chi^{\uparrow \downarrow}(q, 0) & = & \chi_0(q, 0)  +  \eta^2 \left[ \frac {k_F(4 k_F^2 + q^2)}{2^6 \pi^2 q^2} \right.\nn\\
& & \left.- \frac {(4 k_F^2 - q^2)^2}{2^8 \pi^2 q^3}\ln\left| \frac {2k_F + q}{2 k_F - q}\right| \right].\nn
 \end{eqnarray}

The remainder of the calculation is very similar to that for the pairing function contribution. After a final numerical integration, the field renormalization is obtained as
\begin{eqnarray}
& &\langle \langle 
\Theta_{\bf k}^2 
\left[ 1 - 
\partial_{\epsilon_{\bf k}} {\cal R}e \Sigma^\uparrow ({\bf k}, \epsilon_{\bf k} ) 
\right]
\rangle \rangle
\nn
\\
&=&
\lambda
\left[0.398{+}0.199 \eta{+}\left( 0.976{+}0.060 \ln T \right) \eta^2 \right].\nn
\end{eqnarray}


\begin{thebibliography}{90}
\bibitem[{pop()}]{population}
\bibinfo{note}{Where adaptable creatures may out-compete their less flexible
  cousins in fluctuating environments}.
\bibitem[{KB()}]{KB}
\bibinfo{note}{The relationship between the two approaches is along the lines
  of the relationship between density functional theory and Kadanoff-Baym
  theory as discussed in Ref.~\cite{RevModPhys.78.865}}.
\end{thebibliography}
\end{document}